\documentclass[11pt]{article}

\usepackage[]{inputenc}
\usepackage[T1]{fontenc}
\usepackage{hyperref}

\usepackage{color}

\usepackage{fullpage}

\usepackage{mathtools}
\usepackage{empheq}
\usepackage{amssymb}
\usepackage[toc,page]{appendix}

\usepackage{amsthm}
\usepackage{float}
\usepackage{wrapfig}
\usepackage{caption}
\usepackage{graphicx}
\usepackage{pgfplots}
\usepackage{subcaption}
\usepackage{tikz,tikz-3dplot}
\tdplotsetmaincoords{80}{45}
\tdplotsetrotatedcoords{-90}{180}{-90}

\usetikzlibrary{intersections,fadings,decorations.pathreplacing}

  \pgfplotsset{
        compat=1.8}

\theoremstyle{definition}

\title{Dynamical Black Hole Entropy in Effective Field Theory}

\def \be{\begin{equation}}
\def \ee{\end{equation}}
\def \ba{\begin{eqnarray}}
\def \ea{\end{eqnarray}}

\author{Iain Davies and Harvey S. Reall \\
{\small Department of Applied Mathematics and Theoretical Physics, University of Cambridge,} \\ {\small Wilberforce Road, Cambridge CB3 0WA, United Kingdom} \\
{\small id318@cam.ac.uk, hsr1000@cam.ac.uk}
}

\begin{document}
\maketitle

\begin{abstract}
In recent work, Hollands, Kovács and Reall have built on previous work of Wall to provide a definition of dynamical black hole entropy for gravitational effective field theories (EFTs). This entropy satisfies a second law of black hole mechanics to quadratic order in perturbations around a stationary black hole. We determine the explicit form of this entropy for the EFT of 4d vacuum gravity including terms in the action with up to $6$ derivatives. An open question concerns the gauge invariance of this definition of black hole entropy. We show that gauge invariance holds for the EFT of vacuum gravity with up to $6$ derivatives but demonstrate that it can fail when $8$ derivative terms are included. We determine an entropy for Einstein-Gauss-Bonnet theory by treating it as an EFT with vanishing $6$ derivative terms. 

\end{abstract}

\section{Introduction}

The work of Wald in the early 1990s showed that a first law of black hole mechanics holds for any theory of gravity arising from a diffeomorphism invariant Lagrangian \cite{Wald:1993nt}. This work provides a definition of entropy, the Wald entropy, for any stationary (i.e. time independent) black hole solution. It is natural to ask whether a second law of black hole mechanics can also be established for this class of theories. Can one define an entropy for a dynamical (i.e. time-dependent) black hole that (i) depends only on the geometry of a cross-section of the event horizon, (ii) agrees with the Wald entropy in equilibrium, and (iii) increases in any dynamical process?

An early proposal due to Iyer and Wald \cite{Iyer:1994ys} satisfies (i) and (ii) above but the question of whether it satisfies a second law (iii) was left open. 
A suitable definition of entropy has been found for the special case of $f(R)$ theories, i.e. theories with a Lagrangian that is a function of the Ricci scalar. For such theories, the entropy is the integral of $4\pi f'(R)$ over a horizon cross section \cite{Jacobson:1995uq}. This differs from the Iyer-Wald entropy, which suggests that the latter will not satisfy (iii) in general. 

More recently, the second law has been studied perturbatively around a stationary black hole. Wall has found a definition of black hole entropy that satisfies (i) and (ii) and satisfies (iii) to {\it linear order} in perturbation theory \cite{Wall:2015}. Note that the second law implies that the entropy is {\it constant} to linear order (otherwise one could obtain a decrease by reversing the sign of the perturbation). Nevertheless, this is still highly non-trivial e.g. in a general theory the Bekenstein-Hawking entropy (proportional to the horizon area) would not be constant to linear order. Wall's definition involves adding certain terms to the Iyer-Wald entropy so, following \cite{Hollands:2022} we shall refer to it as the Iyer-Wald-Wall (IWW) entropy. 

To see an increase in entropy one needs to go beyond linear perturbation theory. This has been achieved in recent work of Hollands, Kovács and Reall (HKR) \cite{Hollands:2022}. They showed that the IWW entropy can be improved, by adding further terms, to define an entropy that satisfies a second law to {\it quadratic order} in perturbations around a stationary black hole. We shall refer to this improved entropy as the IWWHKR entropy. To establish a second law, HKR introduced two physically reasonable assumptions: (a) the theory must be regarded as an effective field theory (EFT) and (b) the black hole solution considered must lie within the regime of validity of EFT. Assumption (a) means that the Lagrangian is a series of terms with increasing numbers of derivatives, whose coefficients scale as appropriate powers of $l$ where $l$ is a ``UV length scale'' e.g. a $k$-derivative term in the Lagrangian has coefficient proportional to $l^{k-2}$. Point (b) means roughly that if $L$ is any length/time scale associated with a dynamical black hole then this must satisfy $L \gg l$. This assumption seems essential since higher-derivative theories typically admit pathological solutions lying outside the regime of validity of EFT and there is no good reason to expect that such solutions should satisfy a second law. 
HKR established the above results for an EFT describing vacuum gravity (no matter) or gravity coupled to a scalar field. They suggested that similar results should hold for more general EFTs. 

This paper has three aims. The first aim is to study examples for which the HKR terms are non-trivial. The leading EFT corrections to Einstein gravity have $4$ derivatives. However, HKR terms are only generated for theories with $6$ or more derivatives, i.e., the IWWHKR entropy coincides with the IWW entropy for a $4$-derivative theory. To find examples for which the IWWHKR entropy differs from the IWW entropy one needs to consider theories with at least $6$ derivatives. We shall consider the EFT of vacuum gravity in 4d. In vacuum, one can use a field redefinition to eliminate (non-topological) $4$-derivative terms from the action and so the leading EFT corrections to the Einstein-Hilbert Lagrangian have $6$ derivatives. We shall determine the HKR terms in the entropy for this EFT. 

Second, we shall study the gauge-invariance of the HKR method. As with with the method of Wall, the HKR approach makes use of Gaussian null coordinates defined near the black hole horizon. One of these coordinates is an affine parameter along the horizon generators. There is the freedom to rescale this affine parameter by a different amount along each generator. The Wall and HKR approaches do not maintain covariance w.r.t. such a rescaling. Nevertheless, HKR proved that the IWW entropy is suitably gauge invariant under the rescaling. However, they did not demonstrate that the new terms generated by their approach must also be gauge invariant. By classifying the possible form of HKR terms in vacuum gravity we shall show that non-gauge-invariant terms cannot arise for theories with fewer than $8$ derivatives. In particular our results for $6$-derivative theories are gauge invariant. However, our classification shows that non-gauge-invariant terms might appear for theories with $8$ derivatives. By considering a particular $8$-derivative theory we confirm that such terms do appear. Hence the IWWHKR entropy is not gauge invariant for this theory. 

Our final objective is to consider a particularly interesting theory, namely Einstein-Gauss-Bonnet (EGB) theory. In $d>4$ dimensional vacuum gravity, field redefinitions can be used to bring terms with up to $4$-derivatives to the EGB form. Thus EGB can be regarded as an EFT. However, 
since this theory has second order equations of motion, it is often regarded as a self-contained classical theory (e.g. it admits a well-posed initial value problem if the curvature is small enough \cite{Kovacs:2020ywu}). It is interesting to ask whether the HKR approach can be used to determine the entropy for this theory. To do this, we can regard it as a very special EFT for which the coefficients of all terms with more than $4$ derivatives are exactly zero. We can then apply the HKR method to this EFT to determine the entropy to any desired order in the Gauss-Bonnet coupling constant. We shall use this method to calculate the entropy to quadratic order in this coupling constant. 

This paper is organized as follows. In section \ref{sec:review} we review the Wall and HKR algorithms. In section \ref{Implementation} we explain how we calculate the IWWHKR entropy in practice using computer algebra. Section \ref{Classify} presents a classification of terms that can arise in the IWWHKR entropy, explaining why this is necessarily gauge invariant for theories with up to $6$ derivatives. Section \ref{Examples} presents the results of our calculations for EGB theory, $6$-derivative theories, and a particular $8$-derivative theory. In section \ref{GaugeInvariance} we demonstrate gauge non-invariance of the IWWHKR entropy for the $8$-derivative theory. Section \ref{sec:discuss} contains a brief discussion. 

Our conventions are a positive signature metric, Greek indices $\mu,\nu,\ldots$ denote spacetime coordinate indices and $(r,v,x^A)$ refer to the Gaussian Null coordinate system defined below. We use units such that $16 \pi G=1$.

\section{Review of Wall and HKR Procedures}

\label{sec:review}

Let us first review the Wall and HKR procedures for constructing a dynamical black hole entropy that satisfies a second law. The Wall procedure is sketched in \cite{Wall:2015} and described in more detail in \cite{Bhat:2020,Bhattacharyya:2021jhr}. The HKR procedure is detailed in \cite{Hollands:2022}.

\subsection{Gaussian Null Co-ordinates}

Both procedures involve an entropy current defined on the event horizon $\mathcal{N}$ of a black hole, which is a null hypersurface. We assume $\mathcal{N}$ is smooth and has generators that extend to infinite affine parameter to the future. This seems reasonable for a black hole ``settling down to equilibrium'', which is the physical situation considered by Wall and HKR. 

We will use Gaussian Null Co-ordinates (GNCs) defined in a neighbourhood of $\mathcal{N}$ as follows. Assume all generators intersect a spacelike cross-section $C$ exactly once, and take $x^A$ to be a co-dimension 2 co-ordinate chart on $C$. Let the null geodesic generators have affine parameter $v$ and future directed tangent vector $l^\mu$ such that $l = \partial_{v}$ and $v=0$ on $C$. We can transport $C$ along the null geodesic generators a parameter distance $v$ to obtain a foliation $C(v)$ of $\mathcal{N}$. Finally, we uniquely define the null vector field $n^\mu$ by $n \cdot (\partial/\partial x^A) =0$ and $n \cdot l = 1$. The co-ordinates $(r, v, x^A)$ are then assigned to the point affine parameter distance $r$ along the null geodesic starting at the point on $\mathcal{N}$ with coordinates $(v,x^A)$ and with tangent $n^\mu$ there. The metric in these GNCs is given by 
\begin{equation}
    g = 2 \text{d}v( \text{d}r - \frac{1}{2}r^2 \alpha \text{d}v -r\beta_{A} \text{d}x^A) + \mu_{A B} \text{d}x^A \text{d}x^B, \quad l = \partial_{v}, \quad n=\partial_{r}
\end{equation}
$\mathcal{N}$ is the surface $r=0$, and $C$ is the surface $r=v=0$. The inverse of $\mu_{A B}$ is denoted by $\mu^{A B}$, and we raise and lower $A, B, ...$ indices with $\mu^{A B}$ and $\mu_{A B}$. We denote the induced volume form on $C(v)$ by $\epsilon_{A_1 ... A_{d-2}} = \epsilon_{r v A_1 ... A_{d-2}}$ where $d$ is the dimension of the spacetime. The covariant derivative on $C(v)$ with respect to $\mu_{A B}$ is denoted by $D_{A}$. We also define
\begin{equation}
    K_{A B} \equiv \frac{1}{2} \partial_{v}{\mu_{A B}}, \quad \bar{K}_{A B} \equiv \frac{1}{2} \partial_{r}{ \mu_{A B} }, \quad K \equiv K^{A}\,_{A}, \quad \bar{K} \equiv \bar{K}^{A}\,_{A}
\end{equation}
$K_{AB}$ describes the expansion and shear of the horizon generators. $\bar{K}_{AB}$ describes the expansion and shear of the ingoing null geodesics orthogonal to a horizon cut $C(v)$. 

\subsection{Gauge Transformations and Boost Weight}\label{GaugeInvarianceLaws}

Our choice of GNCs is not unique. In particular, on each generator we can rescale the affine parameter $v' = v / a(x^A)$ and tangent vector $l'= a(x^A) l$, where $a(x^A)>0$, to obtain a new set of GNCs $(r', v', x^A)$. If $a(x^A)$ is non-constant then horizon cross-sections of constant $v'$ differ from cross-sections of constant $v$, i.e., the change of coordinates changes the foliation $C(v)$. The exception is the surface $C$, i.e., $v=v'=0$, which belongs to both foliations. 

This freedom in rescaling the affine parameter is essential if we wish to compare the entropy of two arbitrary horizon cuts $C$, $C'$ with $C'$ strictly to the future of $C$ \cite{Hollands:2022}. The reason is that the Wall and HKR procedures define an entropy associated with a particular set of GNCs. Introducing GNCs based on $C$ we can use the rescaling of affine parameter to ensure that $C'$ is a surface of constant $v$, and then apply the second law of Wall or HKR. However, 
we do not want our definition of the entropy of $C$ to depend on the choice of $C'$ and so 
we want the entropy of $C$ to be gauge invariant under the rescaling of affine parameter. Investigating whether or not this is true for the IWWHKR entropy is one of the aims of this paper.

In general, GNC quantities have complicated transformation laws under this rescaling. However, we will only be concerned with how they transform on the horizon cut $C$, i.e. on $r=v=0$. On $C$, some quantities satisfy simple transformation laws \cite{Hollands:2022}:
\begin{align}\label{GNCTransform}
    x'^{A} = x^A, \quad v'=&\, 0 \quad r' = 0,\nonumber\\
    \mu'_{A B} = \mu_{A B}, \quad \epsilon'_{A_1 ... A_{d-2}}= \epsilon_{A_1 ... A_{d-2}}, \quad &R_{A B C D}[\mu'] = R_{A B C D}[\mu], \quad D'_A = D_A,\nonumber\\
    \partial_{v'}^p K'_{A B} = a^{p+1} \partial_{v}^p K_{A B}, \,\,\,\,& \partial_{r'}^p \Bar{K}'_{A B} = a^{-p-1} \partial_{r}^p \Bar{K}_{A B},\\
    \beta'_{A} = \beta_A +& 2 D_{A} \log a\nonumber\\
    \text{on} \,\, v=&r=0\nonumber
\end{align}
Note in particular that $\beta_A$ transforms inhomogeneously, in the same way as a gauge field (this is because $\beta_A$ is a connection on the normal bundle of $C$ \cite{Hollands:2022}).

An important concept for both the IWW and HKR procedures is the \textit{boost weight} of a quantity. Suppose we take $a(x^A)$ to be {\it constant}, and suppose that a quantity $T$ transforms as $T'=a^b T$ under the rescaling above. Then $T$ is said to have boost weight $b$. See \cite{Hollands:2022} for a full definition. Some important facts are stated here:
\begin{itemize}
    \item $\alpha$, $\beta_{A}$, and $\mu_{A B}$ have boost weight 0. $K_{A B}$ and $\bar{K}_{A B}$ have boost weight $+1$ and $-1$ respectively.
    \item  If $T$ has boost weight $b$, then $D_{A_1}{... D_{A_n}{ \partial_{v}^{p} \partial_{r}^q T } }$ has boost weight $b+p-q$.
    \item If $X_{i}$ has boost weight $b_i$ and $T= \prod_{i} X_{i}$, then $T$ has boost weight $b=\sum_{i} b_{i}$.
    \item A tensor component $T^{\mu_1 ... \mu_n}_{\nu_1 ... \nu_m}$ has boost weight given by the sum of $+1$ for each $v$ subscript and each $r$ superscript and $-1$ for each $r$ subscript and $v$ superscript.
\end{itemize}

For non-constant $a(x^A)$, a tensor component $T^{\mu_1 ... \mu_n}_{\nu_1 ... \nu_m}$ with boost weight $b$ has a complicated transformation away from $C$ under the rescaling above. However, on $C$ the transformation remains $T'^{\mu_1 ... \mu_n}_{\nu_1 ... \nu_m} = a^b T^{\mu_1 ... \mu_n}_{\nu_1 ... \nu_m}$

\subsection{Perturbations around Stationary Black Holes}

The Wall and HKR results apply to the scenario of a black hole settling to an equilibrium stationary state. To make this precise, we assume that our higher derivative gravitational theories admit a family $\mathcal{F}$ of stationary black hole solutions for which the event horizon is a bifurcate Killing horizon. This is equivalent to the assumption that a member of $\mathcal{F}$ satisfies the zeroth law of black hole mechanics \cite{Racz:1992bp}. For an EFT describing vacuum gravity with UV length scale $l$, a zeroth law can be proved for solutions constructed as a power series in $l$ \cite{Ghosh:2020dkk,Bhat:2022}.\footnote{ The assumption of an expansion in $l$ seems reasonable for stationary black holes but it would certainly not be appropriate in a time-dependent situation, where secular terms (growing in time) typically arise in such an expansion. A simple example of this would be a quasinormal mode of a linearly perturbed black hole. Such a mode is proportional to $e^{-i\omega t}$ with ${\rm Im}(\omega)<0$. Higher derivative terms will give a perturbative shift in the quasinormal frequency, i.e., $\omega = \omega_0 + l^2 \omega_1 + \ldots$. Low lying quasinormal modes certainly lie within the regime of validity of EFT, and decay exponentially in time. However if we expand in $l$ we obtain secular growth: $e^{-i\omega t} = e^{-i \omega_0 t}(1- i \omega_1 l^2 t + \ldots)$. We emphasize that in this paper we do not require that dynamical black hole solutions are constructed as a power series in $l$.} (In some cases a zeroth law can be established assuming only validity of EFT without resorting to an expansion in $l$ \cite{Reall:2021voz}.) A black hole ``rigidity theorem'', (showing that the event horizon of a stationary black hole must be a Killing horizon) has also been proved for solutions of this type \cite{Hollands:2022ajj}.

The bifurcate Killing horizon assumption ensures that all positive boost weight quantities vanish on $\mathcal{N}$ for a member of $\mathcal{F}$. Therefore, in perturbation theory around a member of $\mathcal{F}$, any positive boost weight quantity is of first order on $\mathcal{N}$. A quantity that is of cubic or higher order in positive boost weight quantities vanishes to quadratic order in perturbations around a stationary black hole. 

\subsection{The Iyer-Wald-Wall Entropy}

\label{sec:IWW}

In standard 2-derivative GR coupled to matter satisfying the null energy condition, the second law of black hole mechanics states that the area of a black hole is non-decreasing, i.e. 
\begin{equation}
    A(v) = \int_{C(v)} \text{d}^{d-2}x \sqrt{\mu}
\end{equation}
satisfies $\dot{A}(v) \ge 0$ for all $v$. The interpretation of the area as the Bekenstein-Hawking entropy re-expresses this law as saying the entropy of a black hole is always non-decreasing.

However, if we consider a higher derivative theory of gravity then there is no reason why $A(v)$ should be non-decreasing for a general black hole solution. Thus, in the hope of preserving our interpretation of black holes as thermodynamic objects, we must come up with a new definition of dynamical black hole entropy that does satisfy a second law.

The Iyer-Wald-Wall entropy is an attempt to do this. It satisfies a second law to linear order in perturbations around a member of $\mathcal{F}$. The starting point for this definition is the $v v$ component of the equations of motion, $E_{v v}$, evaluated on $\mathcal{N}$. In a procedure sketched by Wall in \cite{Wall:2015}\footnote{
See \cite{Chatterjee:2011wj,Bhattacharjee:2015yaa} for earlier work establishing a linearized second law in particular theories.
}  and made explicit by Bhattacharyya {\it et al} \cite{Bhat:2020,Bhattacharyya:2021jhr}, $E_{v v}$ can be manipulated into the following form on $\mathcal{N}$:
\begin{equation} \label{IWW}
    E_{v v}\Big|_{\mathcal{N}} = \partial_{v}\left[\frac{1}{\sqrt{\mu}} \partial_{v}\left(\sqrt{\mu} s^{v}_{IWW}\right) + D_{A}{ s^A }\right] + ...
\end{equation}
where $s^v_{IWW}$ (boost weight $0$) and $s^A$ (boost weight $1$) are expressions containing terms of up to linear order in positive boost weight quantities and the ellipsis denotes terms that are quadratic or higher in positive boost weight quantities (and therefore of quadratic or higher order in perturbation theory around a member of $\mathcal{F}$). The method to get to (\ref{IWW}) is purely off-shell (i.e. it doesn't use any equations of motion) and applies to any theory of gravity arising from a diffeomorphism invariant Lagrangian. 

The IWW entropy of a horizon cross-section $C(v)$ is defined as\footnote{We use units with $16\pi G=1$.}
\begin{equation}
    S_{IWW}(v) = 4 \pi \int_{C(v)} \text{d}^{d-2}x \sqrt{\mu} \,s^{v}_{IWW} 
\end{equation}
We can then write
\begin{equation}
\begin{split}
    \dot{S}_{IWW} =& 4 \pi\int_{C(v)} \text{d}^{d-2}x \sqrt{\mu}\left[\frac{1}{\sqrt{\mu}} \partial_{v}(\sqrt{\mu} s^{v}_{IWW}) + D_{A}{s^A}\right] \\
    =& - 4 \pi \int_{C(v)} \text{d}^{d-2}x \sqrt{\mu} \int_{v}^{\infty} \text{d}v \, \partial_{v} \left[ \frac{1}{\sqrt{\mu}} \partial_{v}(\sqrt{\mu} s^{v}_{IWW}) + D_{A}{s^A} \right]
\end{split}
 \end{equation}
 where in the first line we trivially added the total derivative $\sqrt{\mu} D_{A}{s^A}$ to the integrand, and in the second line we assumed the black hole settles to a member of $\mathcal{F}$, so positive boost weight quantities vanish on the horizon at late times. The integrand can then be swapped for terms that are quadratic or higher in positive boost weight quantities using (\ref{IWW}) and the equation of motion $E_{v v}=0$. Thus $\dot{S}_{IWW}=0$ up to linear order in perturbations around a stationary black hole, and hence satisfies a second law up to this order.
 
$s^{v}_{IWW}$ is the IWW entropy density described by Wall, whilst the need for the quantity $s^A$ was noted in \cite{Bhat:2020}. These objects can be combined into a vector field $(s^v_{IWW},s^A)$ tangent to $\mathcal{N}$, called the {\it entropy current}. How to construct $s^{v}_{IWW}$ and $s^A$ is briefly discussed in Section \ref{Implementation}. $s^{v}_{IWW}$ has boost weight 0 and contains terms that are either linear or zero order in positive boost weight quantities. The zero order terms are exactly the Iyer-Wald entropy density, and hence the IWW entropy density modifies this by terms linear in positive boost weight. 

We have defined $s^v_{IWW}$ so that it includes only terms of up to linear order in positive boost weight quantities. However, one can add terms to $s^v_{IWW}$ of quadratic or higher order in such quantities without affecting the linearized second law. Some previous work on the IWW entropy (e.g. \cite{Wall:2015}) implicitly makes a choice for these higher order terms. We emphasize that our definition of $s^v_{IWW}$ does not include such terms. 

HKR proved that the IWW entropy is gauge invariant under changes of GNCs \cite{Hollands:2022}. More precisely, they showed that it is gauge invariant to linear order in perturbations, i.e., to the same order that it satisfies the second law. They also showed that it can be made gauge invariant to all orders by adding higher order terms as discussed in the previous paragraph. The quantity $s^A$ is not gauge invariant in general \cite{Bhattacharyya:2022njk,Hollands:2022}.

\subsection{Gravitational Effective Field Theory}

At linear order, the second law states that the entropy is constant. To obtain an entropy increase it is necessary to go beyond first order perturbation theory. To do this, HKR worked within the framework of effective field theory (EFT). 

In the EFT framework for gravity, we assume the GR Lagrangian comes with arbitrary corrections made out of any diffeomorphism invariant quantity, ordered by the number of derivatives they contain. The coefficients of these corrections are assumed to scale as appropriate powers of some ``UV length scale'' $l$, with an $n$-derivative term having coefficient proportional to $l^{n-2}$. In vacuum gravity the general EFT Lagrangian is of the form
\begin{equation} \label{EFTLag}
    \mathcal{L} = -2\Lambda + R + l^2 \mathcal{L}_{4} + l^4 \mathcal{L}_{6} + l^6 \mathcal{L}_{8} + O(l^8)
\end{equation}
where $\mathcal{L}_{4}$ contains all independent 4-derivative terms, $\mathcal{L}_{6}$ contains all independent 6-derivative terms, $\mathcal{L}_{8}$ contains all independent 8-derivative terms etc.\footnote{
For even spacetime dimension or a parity-symmetric theory, only even numbers of derivatives appear in ${\cal L}$. In this discussion we assume we are working with such a theory.}

The set of terms in each $\mathcal{L}_n$ can be reduced by neglecting total derivatives and using field redefinitions. For example, let us consider $\mathcal{L}_4$. We can write the most general set of independent 4-derivative terms in the above as:
\begin{equation}
    \mathcal{L} = -2 \Lambda + R + l^2\big( k_{1} R^2 + k_{2} R_{\mu \nu} R^{\mu \nu} + k_{3} R_{\mu \nu \rho \sigma} R^{\mu \nu \rho \sigma} + k_{4} \nabla^{\mu}{ \nabla_{\mu}{ R } } \big) + O(l^4)
\end{equation}
The final 4-derivative term is a total derivative, which we can neglect. We can also use a field redefinition of the form $g_{\mu \nu} \rightarrow g_{\mu \nu} + a g_{\mu \nu} + l^2(b R g_{\mu \nu} + c R_{\mu \nu})$ with appropriate constants $a, b,c$ to bring the remaining 4-derivative terms into the form of a single term given by the EGB Lagrangian \footnote{This process may also renormalize the values of $\Lambda$ and $G$.}: 
\begin{equation}
    \mathcal{L} = -2 \Lambda + R + \frac{1}{16} k l^2 \delta^{\rho_1 \rho_2 \rho_3 \rho_4}_{\sigma_1 \sigma_2 \sigma_3 \sigma_4} R_{\rho_1 \rho_2}\,^{\sigma_1 \sigma_2} R_{\rho_3 \rho_4}\, ^{\sigma_3 \sigma_4} + O(l^4)
\end{equation}
where the generalized Kronecker delta is
\begin{equation}
    \delta^{\rho_1 ... \rho_n}_{\sigma_1 ... \sigma_n}= n! \delta^{\rho_1}_{[\sigma_1} ... \delta^{\rho_n}_{\sigma_n]}
\end{equation}

For general dimension $d>4$, this is as far as we can go. In this case, EGB gravity is the leading order EFT correction to GR. Its IWWHKR entropy is calculated in Section \ref{EGBCalc}.

However, in $d=4$ dimensions the 4-derivative EGB term is topological and can be neglected.\footnote{
This topological term in the action contributes a topological (constant) term to the entropy which does not play a role in the situation we are considering of a black hole settling down to equilibrium.} Therefore we can eliminate all order $l^2$ contributions to $\mathcal{L}$. We can perform a similar procedure \cite{Endlich:2017, Cano:2019} to reduce the set of 6-derivative terms to the following:
\begin{equation}
    \mathcal{L} = -2 \Lambda + R + l^4( k_{1} R_{\mu \nu \kappa \lambda} R^{\kappa \lambda \chi \eta} R_{\chi \eta}\,^{\mu \nu} + k_{2} R_{\mu \nu \kappa \lambda} R^{\kappa \lambda \chi \eta} R_{\chi \eta \rho \sigma} \epsilon^{\mu \nu \rho \sigma} ) + O(l^6)
\end{equation}
The $k_1$ term is parity even and the $k_2$ term is parity odd. The IWWHKR entropy of this Lagrangian is calculated to order $l^4$ in Section \ref{CubicCalc}. The contribution at order $l^6$ of a specific term in $\mathcal{L}_8$ is calculated in Section \ref{QuarticCalc}.

Following HKR, we shall not be interested in arbitrary solutions of these EFTs but only solutions that lie within the regime of validity of EFT. HKR define this criterion as follows. We assume we have a $1$-parameter family of dynamical black hole solutions labelled by a length $L$ (this could be the size of the black hole or some other dynamical length/time scale) such that $\mathcal{N}$ is the event horizon for all members of the family (this is a gauge choice). We assume that there exist GNCs defined near $\mathcal{N}$ such that any quantity constructed from $n$ derivatives of $\{\alpha,\beta_A,\mu_{AB} \}$ is bounded by $C_n/L^n$ for some constant $C_n$, and that $|\Lambda| L^2 \le 1$. Then the solution lies within the regime of validity of EFT for sufficiently small $l/L$. This definition captures the notion of a solution ``varying over a length scale $L$'' with $L$ large compared to the UV scale $l$.

\subsection{HKR Procedure}

A complete description of gravity in the regime of validity of EFT would require knowing the coefficients of all terms for all orders of $l$ in the Lagrangian (\ref{EFTLag}). However in practice this is not possible, and so there we will be an $N$ for we which we only know the terms with $N$ or fewer derivatives. Thus the equation of motion will take the form
\begin{equation}
\label{eofm}
 E_{\mu\nu} = O(l^N)
\end{equation}
where the LHS denotes the known terms with up to $N$ derivatives and the RHS denotes the unkown terms with $N+2$ or more derivatives. On the RHS we should really write $O(l^N/L^{N+2})$ but we shall suppress the $L$-dependence henceforth. This dependence can be reinstated by dimensional analysis.

Since the RHS of the equation of motion is unknown, in EFT one cannot prove a second law that holds exactly. The best one can hope for is to find a definition of entropy that satisfies a second law to the same accuracy as the theory itself, i.e. modulo terms of order $O(l^N)$. HKR showed how to define an entropy $S(v)$ containing terms with up to $N-2$ derivatives that satisfies  a second law to quadratic order in perturbations around a stationary black hole, modulo terms of order $O(l^N)$:
\begin{equation}
    \delta^2 \dot{S}(v) \ge -O(l^N)
\end{equation}
where $\delta^2$ indicates a second order perturbation around a member of $\mathcal{F}$ and the minus sign on the RHS indicates that $\delta^2 \dot{S}(v)$ might be negative but only by a small amount of order $O(l^N)$. 

To do this, HKR showed that the terms that are quadratic or higher order in positive boost weight quantities in (\ref{IWW}) (i.e. the terms denoted by the ellipsis) can be brought into the following form on $\mathcal{N}$:
\begin{multline}\label{HKRForm}
    E_{v v}\Big|_{\mathcal{N}} - \partial_{v}\left[\frac{1}{\sqrt{\mu}} \partial_{v}\left(\sqrt{\mu} s^{v}_{IWW}\right) + D_{A}{ s^A }\right] =\\
    \partial_{v}\left[\frac{1}{\sqrt{\mu}} \partial_{v}\left(\sqrt{\mu} \varsigma^{v}\right)\right] + \left(K_{A B}+X_{A B}\right) \left(K^{A B}+X^{A B}\right) + D_{A}{Y^{A}} + O(l^{N})
\end{multline}
where $\varsigma^{v} = \sum_{n=2}^{N-2}l^n \varsigma_n^v$ (boost weight $0$) and $Y^A=\sum_{n=2}^{N-2}l^n Y_n^A$ (boost weight $2$) are of quadratic or higher order in positive boost weight quantities, and $X^{A B}=\sum_{n=2}^{N-2}l^n X_n^{A B}$ (symmetric, boost weight $1$) is of linear or higher order in such quantities. 
An important difference between this construction and the Wall procedure is that this construction requires going on-shell by swapping some ``non-allowed'' terms for others using the equations of motion $E_{\mu \nu} = O(l^{N})$. Details of how this is done are given in Section \ref{Implementation}.

The IWWHKR entropy density is defined as
\begin{equation}
    S^{v} = s^{v}_{IWW} + \varsigma^{v}
\end{equation}
and the IWWHKR entropy of a horizon cross-section $C(v)$ is defined by
\begin{equation}
    S_{IWWHKR}(v) = 4 \pi \int_{C(v)} \text{d}^{d-2}x \sqrt{\mu}\, S^{v} 
\end{equation}
We can perform a similar calculation to the IWW case to find
\begin{equation}
    \begin{split}
        \dot{S}_{IWWHKR} =& - 4 \pi\int_{C(v)} \text{d}^{d-2}x \sqrt{\mu} \int_{v}^{\infty} \text{d}v \partial_{v} \left[ \frac{1}{\sqrt{\mu}} \partial_{v}(\sqrt{\mu} S^{v}) + D_{A}{s^A} \right]\\
        =& 4 \pi \int_{C(v)} \text{d}^{d-2} x \sqrt{\mu} \int_{v}^{\infty} \text{d}v \left[\left(K_{A B}+X_{A B}\right) \left(K^{A B}+X^{A B}\right) + D_{A}{Y^{A}} + O(l^{N}) \right]
    \end{split}
\end{equation}
The first term in the integrand is a positive definite form, so must be non-negative. The second term $D_{A}{Y^A}$ need not have a definite sign. However, since $Y^A$ is quadratic in positive boost weight quantities, it is of second order in perturbation theory when we expand around a member of $\mathcal{F}$, i.e., $Y^A = \delta^2 Y^A$ plus higher order terms. We then have
\begin{equation}
    \int_{C(v)} \text{d}^{d-2} x \sqrt{\mu} \int_{v}^{\infty} \text{d}v D_{A}{\delta^2 Y^A} = \int_{C(v)} \text{d}^{d-2} x \sqrt{\mu}\bigg|_{\mathcal{F}} \int_{v}^{\infty} \text{d}v D_{A}\bigg|_{\mathcal{F}} \delta^2 Y^A + ...
\end{equation}
where the ellipsis denotes terms of cubic or higher order in perturbation theory. Since $\mu_{A B}$ is independent of $v$ when evaluated on a member of $\mathcal{F}$, we can exchange the order of integrations and see the integrand is a total derivative. Hence the integral vanishes to quadratic order in perturbations around a stationary black hole, and so $\dot{S}_{IWWHKR}$ is non-negative to quadratic order, modulo $O(l^{N})$ terms.

\section{Implementation of Algorithm}\label{Implementation}

We shall now discuss how one calculates the IWWHKR entropy in practice. Since the equations involved get extremely lengthy, a symbolic computer algebra program is needed to do this. The program of choice of the authors is Cadabra\cite{Cadabra1}\cite{Cadabra2}\cite{Cadabra3} due to its ability to split $\mu, \nu, ...$ indices into $r$, $v$ and $A, B, C, ...$ indices, and the ease with which expressions can be canonicalised using symmetry or anti-symmetry of indices.  

We assume that we know the terms with up to $N$ derivatives in the EFT Lagrangian. The Lagrangian can be written $\mathcal{L}+O(l^N)$ where $O(l^N)$ is the contribution from the unknown terms with $N+2$ or more derivatives and  
\begin{equation}\label{HDLag}
    \mathcal{L} = -2\Lambda + R + \mathcal{L}_{higher} 
\end{equation}
where $\mathcal{L}_{higher} = \sum_{n=2}^{N-2} l^{n} \mathcal{L}_{n+2}$ are the known higher derivative terms. We shall break down how to calculate the HKR entropy density for this Lagrangian into steps.

\subsection{Calculate Equations of Motion}
First we calculate the equations of motion for this Lagrangian. This gives \eqref{eofm} with
\begin{equation}
    E_{\mu \nu}\equiv -\frac{1}{\sqrt{-g}} \frac{\delta(\sqrt{-g}\mathcal{L})}{\delta g^{\mu \nu}} = -\Lambda g_{\mu \nu} - R_{\mu \nu} + \frac{1}{2}R g_{\mu \nu} + H_{\mu \nu}
\end{equation}
where $H_{\mu \nu}=\sum_{n=2}^{N-2} l^{n} H_{n\, \mu \nu}$ is the contribution from the known higher derivative terms. 

\subsection{Calculate IWW Entropy Current}
Second we must find the IWW entropy current $s^{v}_{IWW}$ and $s^A$ (of equation \eqref{IWW}) for this theory. Since they are not the main subject of this paper, we will only touch briefly on how to calculate them. Full procedures are given in \cite{Bhattacharyya:2021jhr}, but since the algorithms have many steps there are no simple formulas for general $\mathcal{L}$. However, if the Lagrangian depends only on the Riemann tensor and not its derivatives, then there is a formula for $s^{v}_{IWW}$ calculated by Wall in \cite{Wall:2015}, which reproduces a formula for holographic entanglement entropy derived previously by Dong \cite{Dong:2013qoa}. This formula involves taking partial derivatives of the Lagrangian with respect to Riemann components\footnote{We define $\partial/(\partial R_{\mu \nu \rho \sigma})$ to have the same symmetries as $R_{\mu \nu \rho \sigma}$ and to be normalised such that a first variation of some quantity $X(R_{\mu \nu \rho \sigma})$ will give $\delta X = \delta R_{\mu \nu \rho \sigma} \frac{\partial X}{\partial R_{\mu \nu \rho \sigma}}$.}, and then discarding terms that are quadratic (or higher order) in positive boost weight quantities:\footnote{
Actually, Wall does not discard these terms. As discussed in section \ref{sec:IWW}, one can include these terms without affecting the validity of the linearized second law. But our definition of $s^v_{IWW}$ requires that such terms are discarded. Whether or not one includes these terms does not affect the IWWHKR entropy as we shall discuss in section \ref{CubicCalc}.} 
\begin{equation}\label{WallIWW}
    s^{v}_{IWW} = -2\left( \frac{\partial \mathcal{L}}{\partial R_{r v r v }}\bigg|_{\mathcal{N}} + 4 \frac{\partial^2 \mathcal{L}}{\partial R_{v A v B } \partial R_{r C r D}}\bigg|_{\mathcal{N}} K_{A B} \bar{K}_{C D} \right) - s^{v}_{quadratic}
\end{equation}
Here $s^{v}_{quadratic}$ is purely there to cancel any terms that are quadratic or higher order in positive boost weight quantities when the Riemann components are expanded out in GNCs on $\mathcal{N}$, so that overall $s^{v}_{IWW}$ has only linear or zero order terms. 

We can apply this to our EFT Lagrangian (\ref{HDLag}) if $\mathcal{L}_{higher}$ is a polynomial in $R_{\mu \nu \rho \sigma}$. The contribution from the Einstein-Hilbert Lagrangian is $1$, and so
\begin{equation}\label{WallIWWEFT}
    s^{v}_{IWW} = 1 -2\left( \frac{\partial \mathcal{L}_{higher}}{\partial R_{r v r v }}\bigg|_{\mathcal{N}} + 4 \frac{\partial^2 \mathcal{L}_{higher}}{\partial R_{v A v B } \partial R_{r C r D}}\bigg|_{\mathcal{N}} K_{A B} \bar{K}_{C D} \right) - s^{v}_{quadratic} 
\end{equation}
None of the Lagrangians we consider in Section \ref{Examples} contain derivatives of Riemann components, and hence (\ref{WallIWWEFT}) is the formula we use to calculate $s^{v}_{IWW}$. 

There is no simple formula for $s^{A}$ in general, and so we must follow the procedure in \cite{Bhattacharyya:2021jhr} to calculate it. The method requires finding the total derivative term $\Theta^{\mu}$ given by $\delta (\sqrt{-g} \mathcal{L}) = \sqrt{-g} (E^{\mu \nu} \delta g_{\mu \nu} + D_{\mu} \Theta^{\mu}[\delta g])$, and the Noether charge for diffeomorphisms $Q^{\mu \nu}$ given by $\nabla_{\nu}{ Q^{\mu \nu}} = 2 E^{\mu \nu} \zeta_{\nu} + \Theta^{\mu}[\partial \zeta] - \zeta^{\mu} \mathcal{L} $ where we have set $\delta g_{\mu \nu} = \nabla_{\mu}{ \zeta_{\nu} } + \nabla_{\nu}{ \zeta_{\mu}}$ in the argument of $\Theta^{\mu}$. For any given theory, both of these quantities can be calculated via theorems given in \cite{Bhattacharyya:2021jhr}.

\subsection{Calculate Remaining Terms in $E_{v v}$}
Now that we have $E_{v v}$, $s^{v}_{IWW}$ and $s^A$, we proceed by expanding them out in GNCs\footnote{Expanding out $E_{v v}$ in GNCs requires expanding out Riemann components and possibly their covariant derivatives in GNCs. The appendix of \cite{Hollands:2022} lists the expansions of all Riemann components, which is sufficient for the EGB Lagrangian calculation in Section \ref{Examples}. However, the cubic and quartic Lagrangians below have equations of motion that depend on $\nabla_{\mu}{ R_{\nu \rho \sigma \kappa} }$ and $\nabla_{\mu}{ \nabla_{\nu}{ R_{\rho \sigma \kappa \lambda} } }$, and so these must also be calculated in terms of GNCs. Beware, even for Cadabra this is an extensive computation.} on $\mathcal{N}$ and calculating the terms denoted by the ellipsis in equation (\ref{IWW}), which we shall denote $F$:
\begin{equation}
    F\equiv E_{v v}\Big|_{\mathcal{N}} - \partial_{v}\left[\frac{1}{\sqrt{\mu}} \partial_{v}\left(\sqrt{\mu} s^{v}_{IWW}\right) + D_{A}{ s^A }\right]
\end{equation}
By construction, these terms will be of quadratic or higher order in positive boost weight quantities. For example, in standard GR (i.e. $N=2$) we have $s^{v}_{IWW}=1$, $s^A=0$, and $E_{v v}\Big|_{\mathcal{N}} = \mu^{A B} \partial_{v}{K_{A B}}-K_{A B} K^{A B} $. We can use $\partial_{v}{\sqrt{\mu}} = \sqrt{\mu} K$ to obtain $F = K_{A B} K^{A B}$ for standard GR.

Let us consider what possible GNC quantities $F$ can depend on. $F$ is evaluated on $\mathcal{N}$ which is $r=0$, so there is no explicit $r$-dependence. It is also a scalar with respect to $A, B,...$ indices. Therefore $F$ is a polynomial in the remaining quantities we can make out of the GNC metric that are covariant in $A, B,...$ indices: $\alpha, \beta_{A}, \mu_{A B}, \mu^{A B}, \epsilon_{A_1 ... A_{d-2}}, R_{A B C D}[\mu]$ and their $\partial_{v}, \partial_{r}$ and $D_{A}$ derivatives.
Here $\epsilon_{A_1 ... A_{d-2}}$ and $R_{A B C D}[\mu]$ are the induced volume form and induced Riemann tensor\footnote{In $d=4$ dimensions, $C(v)$ is 2-dimensional which implies that $R_{A B C D}[\mu]= \frac{R[\mu]}{2} (\mu_{A C} \mu_{B D} - \mu_{A D} \mu_{B C})$.} on $C(v)$. We can eliminate their $\partial_{v}$ and $\partial_{r}$ derivatives in exchange for $K_{A B}$ and $\Bar{K}_{A B}$ via the formulae
\begin{equation}
    \begin{split}
        \partial_{v}{\epsilon_{A_1 ... A_{d-2}}} =& \epsilon_{A_1 ... A_{d-2}} K,\\
        \partial_{r}{\epsilon_{A_1 ... A_{d-2}}} =& \epsilon_{A_1 ... A_{d-2}} \Bar{K},\\
        \partial_{v}{ R_{A B C D}[\mu]} =& K^{E}\,_{B} R_{A E C D}[\mu] - K^{E}\,_{A} R_{B E C D}[\mu] +D_{C}{D_{B}{K_{A D}}}-\\
        &D_{C}{D_{A}{K_{B D}}}-D_{D}{D_{B}{K_{A C}}}+D_{D}{D_{A}{K_{B C}}},\\
        \partial_{r}{ R_{A B C D}[\mu]} =& \Bar{K}^{E}\,_{B} R_{A E C D}[\mu] - \Bar{K}^{E}\,_{A} R_{B E C D}[\mu] +D_{C}{D_{B}{\Bar{K}_{A D}}}-\\
        &D_{C}{D_{A}{\Bar{K}_{B D}}}-D_{D}{D_{B}{\Bar{K}_{A C}}}+D_{D}{D_{A}{\Bar{K}_{B C}}}
    \end{split}
\end{equation}
Given a term of the form $\partial_{v}^{p} \partial_{r}^q D_{A_1}{... D_{A_n}{ \varphi } }$, we can commute the $D$ derivatives to the left using commutation rules:
\begin{equation}
\begin{split}
    [\partial_{v}, D_{A}]t_{B_1 ... B_n} =& \sum_{i=1}^{n} \mu^{C D} ( D_{D}{ K_{A B_i} } -D_{A}{ K_{D B_i} } - D_{B_i}{ K_{A D} } ) t_{B_1... B_{i-1} C B_{i+1} ... B_n},\\
    [\partial_{r}, D_{A}]t_{B_1 ... B_n} =& \sum_{i=1}^{n} \mu^{C D} ( D_{D}{ \Bar{K}_{A B_i} } -D_{A}{ \Bar{K}_{D B_i} } - D_{B_i}{ \Bar{K}_{A D} } ) t_{B_1... B_{i-1} C B_{i+1} ... B_n}
\end{split}
\end{equation}
$[\partial_{v}, D]$ commutators introduce $D_{A}{K_{B C}}$ terms, whilst $[\partial_{r}, D]$ commutators introduce $D_{A}{ \bar{K}_{B C}}$ terms.

Using these results we can express $F$ as a polynomial in the following quantities:
\begin{equation}\label{GeneralTerms}
    \mu_{A B}, \,\, \mu^{A B}, \,\, \epsilon_{A_1 ... A_{d-2}}, \,\,\, D_{A_1}{... D_{A_n}{ R_{A B C D}[\mu] } } \,\,\, \\
    \text{or} \,\,\, D_{A_1}{... D_{A_n}{ \partial_{v}^{p} \partial_{r}^q \varphi } } \,\,\, \text{for} \,\,\, \varphi \in \{\alpha, \beta_{A}, K_{A B}, \bar{K}_{A B}\}
\end{equation}

\subsection{Reduce to Allowed Terms}

The key step in the HKR algorithm is to reduce the above set of terms to a much smaller set of ``allowed'' terms:
\begin{empheq}[box=\fbox]{align}\label{AllowedTerms}
    \text{Allowed terms:} \,\, \mu_{A B}, \,\, &\mu^{A B}, \,\, \epsilon_{A_1 ... A_{d-2}}, \,\, D_{A_1}{... D_{A_n}{ R_{A B C D}[\mu] } },\nonumber\\
    D_{A_1}{... D_{A_n}{ \beta_{A} }}, \,\, D_{A_1}&... D_{A_n}{\partial_{v}^p K_{A B} }, \,\, D_{A_1}{... D_{A_n}{\partial_{r}^q \bar{K}} }, \,\, \Lambda
\end{empheq}
In particular, the only positive boost weight allowed terms are of the form $D^k\partial_{v}^p K_{A B}$. This reduction is achieved through careful inspection of the GNC expressions for Ricci components and application of the equations of motion. Let us see how this works via an example. 

Consider the GNC expression for $R_{v A}$ on $\mathcal{N}$:
\begin{equation}
    R_{v A}\Big|_{\mathcal{N}} = \frac{1}{2}\partial_{v}{\beta_{A}} + D_{B}{K_{A C}} \mu^{B C}-D_{A}{K_{B C}} \mu^{B C}+\frac{1}{2}K_{B C} \beta_{A} \mu^{B C}
\end{equation}
We can rearrange this to get $\partial_{v}{ \beta_A }$ in terms of allowed terms and the Ricci component $R_{v A}$:
\begin{equation} \label{dvbeta}
    \partial_{v}{\beta_{A}} = - 2 D_{B}{K_{A C}} \mu^{B C}+2 D_{A}{K_{B C}} \mu^{B C}-K_{B C} \beta_{A} \mu^{B C} + 2 R_{v A}\Big|_{\mathcal{N}}
\end{equation}
We can rewrite the equation of motion \eqref{eofm} as $R_{\mu\nu} = \frac{2}{d-2} \Lambda g_{\mu \nu} - \frac{1}{d-2} g^{\rho \sigma} H_{\rho \sigma} g_{\mu \nu} + H_{\mu \nu} + O(l^N)$. Since $H_{\mu \nu}$ is at least $O(l^2)$, we can therefore swap $R_{\mu \nu}$ for $\frac{2}{d-2} \Lambda g_{\mu \nu}$ plus higher order terms in $l$. Thus we can use (\ref{dvbeta}) to eliminate $\partial_v \beta_A$ in favour of allowed terms plus terms of higher order in $l$. Working order by order in $l$ we can therefore eliminate occurrences of $\partial_{v}{ \beta_A }$, pushing them to higher order at each step. Eventually we reach $O(l^N)$, at which point we stop since we do not know the terms in the equation of motion at this order. 

We can find a similar expression for $\partial_{r}{ \beta_A }$ by considering $R_{r A}\Big|_{\mathcal{N}}$:
\begin{equation}
    \partial_{r}{ \beta_{A} } = D_{B}{\bar{K}_{A C}} \mu^{B C}-D_{A}{\bar{K}_{B C}} \mu^{B C}+\bar{K}_{A B} \beta_{C} \mu^{B C} - \frac{1}{2}\bar{K}_{B C} \beta_{A} \mu^{B C} - R_{r A}\Big|_{\mathcal{N}}
\end{equation}
Again we can use the equations of motion to swap out $R_{r A}$ and push any occurrence of $\partial_{r}{ \beta_A }$ to higher order in $l$, eventually reaching $O(l^N)$. 

We can similarly eliminate $\partial_{v}{ \Bar{K}_{A B} }$ using $R_{A B}$, and eliminate $\alpha$ using $R_{v r}$. We can take $D$ and $\partial_{v}$ derivatives of these expressions to eliminate further terms. We can't take $\partial_{r}$ derivatives since the expressions are evaluated on $r=0$, so instead we look at the $\nabla_{r}$ derivatives of the Ricci tensor. For example, to eliminate $\partial_{r}{\alpha}$, we look at $\nabla_{r}{ R_{v r} }$.

When looking at a specific theory, one will only need to calculate a finite number of these elimination rules, as there will be only so many derivatives involved. The set we need for the EGB, cubic and quartic Lagrangians in Section \ref{Examples} are given in Appendix \ref{app:elimination}.

\subsection{Manipulate Terms Order-by-Order}

Once we have eliminated non-allowed terms, we proceed order-by-order in $l$. We can separate terms in $F$: 
\begin{equation}
    F = F_{0} + \sum_{n=2}^{N-2} l^n F_{n} + O(l^N)
\end{equation}
Each $F_n$ will be quadratic in positive boost weight terms and only depend on allowed terms by construction. $F_0$ is the contribution from the 2-derivative GR Lagrangian calculated previously: $F_{0}=K_{A B} K^{A B}$. Note $F_0$ is of the required form (\ref{HKRForm}) at order $l^0$ with $\varsigma_0^v = X_0^{A B} = Y_{0}^A =0$. Hence we work inductively: let us assume we have manipulated $F$ into the correct form up to some order $l^{m-2}$, $m<N$: 
\begin{multline}\label{Fexpanded}
    F= \left(K_{A B}+\sum_{n=2}^{m-2} l^n X_{n A B}\right) \left(K^{A B}+\sum_{n=2}^{m-2} l^n X^{A B}_n\right) +\\
    \partial_{v}\left[\frac{1}{\sqrt{\mu}} \partial_{v}\left(\sqrt{\mu} \sum_{n=2}^{m-2} l^n \varsigma^{v}_n \right)\right] + D_{A}{\sum_{n=2}^{m-2} l^n Y_n^{A}} + \sum_{n=m}^{N-2} l^n F_{n} + O(l^{N})
\end{multline}
with the remaining $F_n$ quadratic in positive boost weight terms and only containing allowed terms. Let us study $F_m$. As noted above, the only positive boost weight allowed terms are of the form $D^k \partial_{v}^p K_{A B}$. Hence each monomial in $F_m$ must have at least two factors of this form, and schematically $F_m$ takes the form
\begin{equation}
    F_m = \sum_{k_1, k_2, p_1, p_2} (D^{k_1}{\partial_{v}^{p_1} K}) \, (D^{k_2}{\partial_{v}^{p_2}K}) \, A_{k_1, k_2, p_1, p_2} 
\end{equation}
where the $A_{k_1, k_2, p_1, p_2}$ (boost weight $-p_1-p_2$) are made up of allowed terms. Note that the $K$'s (and $D$'s) above should have indices $K_{A B}$ but these have been dropped in the schematic form for notational ease. They should not be interpreted as $K\equiv K^{A}\,_{A}$. The $A_{k_1, k_2, p_1, p_2}$ can in principle include more $D^{k}{\partial_{v}^{p} K }$ terms, so for simplicity when performing the algorithm we always order the terms in the following priority: $p_1$ and $p_2$ as small as possible, $p_1\leq p_2$, and then $k_1$ and $k_2$ as small as possible.

We now aim to manipulate this sum so that everything in it is proportional to $K_{A B}$. First we move over the $D^{k_1}$ derivatives in the first factor of each term in the sum using the product rule $D(f) g = - f D(g) + D(f g)$. This will produce some total derivative $D_{A} Y_{m}^A$:
\begin{equation}
    F_m = \sum_{k, p_1, p_2} (\partial_{v}^{p_1} K) \, (D^{k}{\partial_{v}^{p_2}K}) \, A_{k, p_1, p_2} + D_{A}{ Y_m^A } 
\end{equation}
Secondly we move over the $\partial_{v}^{p_1}$ derivatives in each term, We do this in such a way as to produce terms of the form $\partial_{v}\left(\frac{1}{\sqrt{\mu}} \partial_{v}\left(\sqrt{\mu} \varsigma^{v}\right)\right)$, where $\varsigma^v$ will contribute to the IWWHKR entropy density. It is proved in \cite{Hollands:2022} that for any $k\geq0$ and $p_1, p_2\geq 1$, there exist unique numbers $a_j$ such that
\begin{equation}
    (\partial_{v}^{p_1} K) \, (D^{k}{\partial_{v}^{p_2}K}) A_{k, p_1, p_2} = \partial_{v}\left\{\frac{1}{\sqrt{\mu}} \partial_{v}\left[\sqrt{\mu} \sum_{j=1}^{p_1+p_2-1} a_j (\partial_{v}^{p_1+p_2-1-j} K) \, (D^{k}{\partial_{v}^{j-1}K }) A_{k, p_1, p_2} \right]\right\} + ...
\end{equation}
where the ellipsis denotes terms of the form $(\partial_{v}^{\Bar{p}_1} K) \, (D^{\bar{k}}{\partial_{v}^{\Bar{p}_2}K}) A_{\Bar{k}, \Bar{p}_1, \Bar{p}_2}$ with $\Bar{p}_1+\Bar{p}_2 < p_1+p_2$ (which can be dealt with inductively) or $\Bar{p}_1=0$ (which are proportional to $K_{A B}$) or $\Bar{p}_2=0$ (which are proportional to $D^{\bar{k}} K_{A B}$ and so can be made proportional to $K_{A B}$ by moving over the $D$ derivatives and adding to $Y_{m}^A$ as above). How to calculate the $a_j$ is described in Appendix \ref{app:aj}. The new $A_{\Bar{k}, \Bar{p}_1, \Bar{p}_2}$ include terms like $\partial_{v}{ A_{k, p_1, p_2} }$ which will involve non-allowed terms. These must be swapped out using the elimination rules and equations of motion, and so will generate $O(l^2)$ terms.

Thus we can repeat this procedure until we have manipulated $F_m$ into the form
\begin{equation}
    F_m = 2 K_{A B} X^{A B}_m + \partial_{v}\left[\frac{1}{\sqrt{\mu}} \partial_{v}\left(\sqrt{\mu} \varsigma_m^{v}\right)\right] + D_{A} Y_m^A + O(l^2)
\end{equation}
We take $X^{A B}_m$ to be symmetric. It will be linear in positive boost weight quantities, and $\varsigma^{v}_m$ and $Y_{m}^{A}$ will be quadratic in positive boost weight quantities. The $O(l^2)$ terms are also quadratic in positive boost weight quantities, and so can be incorporated into $\sum_{n=m+2}^{N-2} l^n F_{n}$.

Substituting this into (\ref{Fexpanded}) gives
\begin{multline}
    F= \left(K_{A B}+\sum_{n=2}^{m-2} l^n X_{n A B}\right) \left(K^{A B}+\sum_{n=2}^{m-2} l^n X^{A B}_n\right) +2l^m K_{A B} X^{A B}_m+\\
    \partial_{v}\left[\frac{1}{\sqrt{\mu}} \partial_{v}\left(\sqrt{\mu} \sum_{n=2}^{m} l^n \varsigma^{v}_n \right)\right] + D_{A}{\sum_{n=2}^{m} l^n Y_n^{A}} + \sum_{n=m+2}^{N-2} l^n F_{n} + O(l^{N})
\end{multline}

We then complete the square with the first two terms, generating more higher order quadratic terms which get incorporated into $\sum_{n=m+2}^{N-2} l^n F_{n}$. This leaves the desired form of $F$ up to order $l^{m}$ and completes the induction. We perform this procedure through each order of $l$ until all terms have been dealt with up to $O(l^N)$:
\begin{multline}
    F= \left(K_{A B}+\sum_{n=2}^{N-2} l^n X_{n A B}\right) \left(K^{A B}+\sum_{n=2}^{N-2} l^n X^{A B}_n\right) +\\
    \partial_{v}\left[\frac{1}{\sqrt{\mu}} \partial_{v}\left(\sqrt{\mu} \sum_{n=2}^{N-2} l^n \varsigma^{v}_n \right)\right] + D_{A}{\sum_{n=2}^{N-2} l^n Y_n^{A}} + O(l^{N})
\end{multline}
The IWWHKR entropy density is then defined as
\begin{equation}
\label{IWWHKRdef}
    S^{v} = s_{IWW}^{v} + \sum_{n=2}^{N-2} l^n \varsigma_{n}^{v}
\end{equation}

\section{Classification of Possible Terms in the IWWHKR Entropy Density}

\label{Classify}

Before we display calculations of the HKR entropy density for specific Lagrangians, we discuss what possible terms can appear in $l^n \varsigma_{n}^{v}$, which is the addition to $s_{IWW}^{v}$ at order $l^n$. 

We will be interested in the number of derivatives associated with a quantity on $\mathcal{N}$. We can write GNC quantities on $\mathcal{N}$ as derivatives of the metric: $\alpha = -\frac{1}{2} \partial^2_{r}{ g_{v v}}\big|_{\mathcal{N}}$, $\beta_{A} = -\partial_{r}{g_{v A}}\big|_{\mathcal{N}}$ and $\mu_{A B} = g_{A B}$. Hence $\alpha, \beta_{A}, \mu_{A B}$ are associated with 2, 1 and 0 derivatives respectively. This motivates a definition of ``dimension'' to count derivatives from \cite{Hollands:2022}:\\

\textbf{Definition}: \textit{The “dimension” of $\alpha$, $\beta_A$ and $\mu_{A B}$ are 2, 1, 0 respectively. Taking a derivative w.r.t. $v$, $r$ or $x_A$ increases the dimension by 1. Dimension is additive under products.}\\

Note $K_{A B}$ and $\Bar{K}_{A B}$ both have dimension 1. $\epsilon_{A_1 ... A_{d-2}}$ has dimension 0. We also define the dimension of $l$ and $\Lambda$ to be $-1$ and $+2$ respectively so that $E_{\mu \nu} = -\Lambda g_{\mu \nu} - R_{\mu \nu} + \frac{1}{2}R g_{\mu \nu} + \sum_{n} l^n H_{n \mu \nu}$ has consistent dimension 2 on $\mathcal{N}$. This also means the elimination rules for non-allowed terms are dimensionally consistent. A quantity with boost weight $b$ must involve at least $|b|$ derivatives, and so has dimension at least $|b|$.

Now, $\varsigma_{n}^{v}$ arises from manipulating $F_n$, which in turn comes from varying the $(n+2)$-derivative Lagrangian $\mathcal{L}_{n+2}$. Thus $F_n$ has dimension $n+2$. $\varsigma^{v}_{n}$ appears with two extra derivatives in $F_n$:
\begin{equation}
    F_n=\partial_{v}\left[\frac{1}{\sqrt{\mu}} \partial_{v}\left(\sqrt{\mu} \varsigma_n^{v}\right)\right] + ...
\end{equation}
Therefore $\varsigma^{v}_n$ has dimension $n$. It is also boost weight 0 since $F_n$ is boost weight $+2$.

By construction $\varsigma_n^v$ is a sum of terms of the form $(\partial_{v}^{p} K) \, (D^{k}{\partial_{v}^{p'}K }) A_{n, k, p, p'}$ with $k,p,p'\geq0$ and where $A_{n, k, p, p'}$ is made exclusively out of allowed terms. Suppose $A_{n, k, p, p'}$ has dimension $d_{n, k, p, p'}$ and boost weight $b_{n, k, p, p'}$. To match the dimension and boost weight of $\varsigma_n^v$, we have two conditions:
\begin{equation}
    \begin{split}
        d_{n, k, p, p'} &= n-2-p-p'-k \\
        b_{n, k, p, p'} &= -2-p-p'
    \end{split}
\end{equation}
But $d_{n, k, p, p'}\geq|b_{n, k, p, p'}|$ which we can rearrange to give
\begin{equation}
    2p+2p'+k\leq n-4 
\end{equation}

If $n=2$ then the RHS is negative, which is impossible for $k,p,p'\geq0$. Thus $\varsigma_{2}^{v}=0$, and so $S^v$ necessarily agrees with $s_{IWW}^v$ at order $l^2$ \cite{Hollands:2022}.

If $n=4$ then the RHS is 0, and so we must have $k=p=p'=0$. Hence $\varsigma_{4}^v$ must be a sum of terms of the form $K_{A B} K_{C E} A^{A B C E}_{4,0,0,0}$. We have $d_{4, 0, 0, 0}=-b_{4, 0, 0, 0}=2$ so $A^{A B C E}_{4, 0, 0, 0}$ must contain two $r$ derivatives and no other derivatives. The only combinations of allowed terms that have this are $\Bar{K}_{F G} \bar{K}_{H I} T^{A B C E F G H I}$ and $\partial_{r}\bar{K}_{F G} T^{A B C E F G}$, where $T^{A B C E F G H I}$ and $T^{A B C E F G}$ are any combination of the 0-dimension quantities $\mu^{A B}$ and $\epsilon^{A_1 ... A_{d-2}}$. In other words, $\varsigma_{4}^v$ is a sum of terms of the form $K_{A B} K_{C E} \Bar{K}_{F G} \bar{K}_{H I}$ or $K_{A B} K_{C E} \partial_{r}\Bar{K}_{F G}$ with their indices completely contracted in some way with $\mu^{A B}$ or $\epsilon^{A_1 ... A_{d-2}}$. The transformation laws (\ref{GNCTransform}) imply that such terms are gauge invariant on $C$, so the IWWHKR entropy density is gauge invariant up to and including order $l^4$ terms, i.e., it is gauge invariant for a Lagrangian with up to $6$ derivatives. 

If $n\geq 6$ then the above conditions don't rule out gauge non-invariant terms appearing in $\varsigma_n^v$. For example the term $K K \Bar{K} \bar{K} \beta_{A} \beta^{A}$ has dimension 6, boost weight 0 and is quadratic in positive boost weight quantities so might appear in $\varsigma_{6}^v$. But due to the occurrence of $\beta_{A}$, which transforms inhomogeneously, this term is not gauge invariant on $C$.

The above discussion focuses solely on vacuum gravity EFTs. However, \cite{Hollands:2022} also proved that scalar-tensor EFTs (i.e. the EFT of a metric coupled to a scalar field $\phi$) have a corresponding IWWHKR entropy that satisfies a second law to the same order. We can perform a similar classification of terms for this case. The procedure is very similar to the above. There are additional terms of the form $D_{A_1}... D_{A_n} \partial_{v}^p \partial_{r}^q \phi$ that can appear in $F$. We can eliminate terms with both $p, q\geq 1$ via the equation of motion for $\phi$, in a similar fashion to eliminating non-allowed GNC terms. We must therefore add to the set of allowed terms $D_{A_1}... D_{A_n} \partial_{v}^p \phi$ and $D_{A_1}... D_{A_n} \partial_{r}^q \phi$. The scalar field has dimension 0 and boost weight $0$ so $D_{A_1}... D_{A_n} \partial_{v}^p \phi$ is a new allowed term with positive boost weight. The classification above can be repeated with $K_{A B}$ replaced with $\varphi \in \{K_{A B}, \partial_{v} \phi\}$ and $\Bar{K}_{A B}$ replaced with $\Bar{\varphi} \in \{\Bar{K}_{A B}, \partial_{r} \phi\}$. We again find that $\varsigma_2^v=0$ and that $\varsigma_4^v$ is gauge invariant on $C$.

\section{Examples of IWWHKR Entropy Density}\label{Examples}

\subsection{Einstein-Gauss-Bonnet}\label{EGBCalc}

For our first example, we shall consider Einstein-Gauss-Bonnet (EGB) theory. Recall that this describes the leading $4$-derivative EFT corrections to Einstein gravity in $d>4$ dimensions. Specifically, we argued that using field redefinitions and dropping total derivatives, the Lagrangian for the EFT for vacuum gravity can be brought to the form $\mathcal{L}_{EGB}+O(l^4)$ where
\begin{equation}
    \mathcal{L}_{EGB} = -2 \Lambda + R + \frac{1}{16} k l^2 \delta^{\rho_1 \rho_2 \rho_3 \rho_4}_{\sigma_1 \sigma_2 \sigma_3 \sigma_4} R_{\rho_1 \rho_2}\,^{\sigma_1 \sigma_2} R_{\rho_3 \rho_4}\, ^{\sigma_3 \sigma_4}
\end{equation}
The equation of motion is \eqref{eofm} with $N=4$ and
\begin{equation}
    E_{\mu \nu} = - \Lambda g_{\mu \nu} - G_{\mu \nu} + \frac{1}{32} k l^2 g_{\mu \tau} \delta^{\tau \rho_1 \rho_2 \rho_3 \rho_4}_{\nu \sigma_1 \sigma_2 \sigma_3 \sigma_4} R_{\rho_1 \rho_2}\,^{\sigma_1 \sigma_2} R_{\rho_3 \rho_4}\, ^{\sigma_3 \sigma_4} 
\end{equation}
If we view this theory as an EFT with $N=4$ then, as explained above, the IWWHKR entropy coincides with the IWW entropy (as for any $N=4$ theory). However, EGB has the special property of possessing second order equations of motion, despite arising from a 4-derivative Lagrangian. This property means EGB gravity can be considered as a self-contained classical theory rather than just the $N=4$ truncation of an EFT. The IWWHKR procedure can be used to define an entropy for this classical theory as an expansion in $l^2$. The idea is to treat this EGB theory as a very special EFT for which the coefficients of the terms with more than $4$ derivatives are exactly zero (i.e. the Lagrangian is exactly $\mathcal{L}_{EGB}$) and use the IWWHKR procedure to define an entropy order by order in $l^2$. We shall show explicity how this works to order $l^4$. 

Let us proceed with the HKR procedure for this theory. We have calculated the equations of motion above. The next step is to find the IWW entropy current. It is given in \cite{Bhattacharyya:2021jhr} with a different normalisation. In our units it is\footnote{
For a stationary black hole, the IWW (or IWWHKR) entropy reduces to the EGB entropy defined in Ref. \cite{Jacobson:1993xs} (which is reproduced by the method of \cite{Wald:1993nt}). For EGB, the IWW entropy density involves only terms of vanishing boost weight and so, for a non-stationary EGB black hole, the IWW entropy is the same as the Iyer-Wald entropy.
} 
\begin{equation}
    s^{v}_{IWW} = 1 + \frac{1}{2} k l^2 R[\mu], \quad \quad \quad \quad s^{A} = k {l}^{2} \left(D^{A}{K}-D^{B}{K^{A}\,_{B}}\right),
\end{equation}
We then calculate the remaining quadratic terms in $E_{v v}$: 
\begin{equation}
\begin{split}
    F \equiv& E_{v v} - \partial_{v}\left[\frac{1}{\sqrt{\mu}} \partial_{v}\left(\sqrt{\mu} s^{v}_{IWW}\right) + s^A\right]\\
    =& K_{A B} K^{A B} + k {l}^{2} \Big[D^{A}{K} D_{A}{K}-2D_{A}{K} D^{B}{K_{B}\,^{A}}+D^{A}{K_{A B}} D^{C}{K_{C}\,^{B}}-D^{A}{K^{B C}} D_{A}{K_{B C}}+\\
    &D^{A}{K^{B C}} D_{B}{K_{A C}} + K_{A B}\Big( \partial_{v}{K_{C E}}\bar{K}^{A B} \mu^{C E}- \partial_{v}{K_{C E}} \bar{K}\mu^{A B} \mu^{C E}-2 \partial_{v}{K_{C E}}\bar{K}^{B C} \mu^{A E}+\\
    & \partial_{v}{K_{C E}} \bar{K}\mu^{A C} \mu^{B E}+ \partial_{v}{K_{C E}} \bar{K}^{C E}\mu^{A B}+D^{A}{K} \beta^{B}-D^{C}{K} \beta_{C} \mu^{A B}-D^{C}{K^{A}\,_{C}} \beta^{B}+\\
    &D^{C}{K_{C}\,^{E}} \beta_{E} \mu^{A B}+D^{C}{K^{A B}} \beta_{C}-D^{A}{K^{B C}} \beta_{C}+D^{A}{D^{C}{K^{B}\,_{C}}}-D^{A}{D^{B}{K}}-D^{C}{D_{C}{K^{A B}}}+\\
    &D^{C}{D^{A}{K^{B}\,_{C}}}+D^{C}{D_{C}{K}} \mu^{A B}-D^{C}{D^{E}{K_{C E}}} \mu^{A B}+\frac{1}{2}K^{A B} R[\mu]-K^{A B} K^{C E} \bar{K}_{C E}-\\
    &2K^{A C} R^{B}\,_{C}[\mu] - \frac{1}{2}K^{C E} \beta_{C} \beta_{E} \mu^{A B}+\frac{1}{4}K \beta^{C} \beta_{C} \mu^{A B}+\frac{1}{2}K^{A C} \beta^{B} \beta_{C} - \frac{1}{4}K^{A B} \beta^{C} \beta_{C}-\\
    &K^{C E} R^{A}\,_{C}\,^{B}\,_{E}[\mu]+K^{C E} R_{C E}[\mu] \mu^{A B}+K^{C E} K_{C E} \bar{K} \mu^{A B}+2K^{A C} K^{B E} \bar{K}_{C E}-\\
    &K^{A C} K^{B}\,_{C} \bar{K}-K^{C E} K_{C}\,^{F} \bar{K}_{E F} \mu^{A B} \Big) \Big]
\end{split}
\end{equation}
It happens that no non-allowed terms appear in $F$ for this Lagrangian, so we do not yet need to use equations of motion to eliminate such terms. We proceed to move one of the $D$ derivatives over in the terms of the form $(DK) (DK)$, producing a total derivative $D_A(l^2 Y^A_2)$. We find that all remaining terms are proportional to $K_{A B}$, and we can express $F$ in the desired form
\begin{equation} \label{EGBl2}
    F = \left(K_{A B}+l^2 X_{2 A B}\right) \left(K^{A B}+l^2 X_{2}^{A B}\right) + D_{A}(l^2 Y_{2}^{A}) - l^4 X_{2 A B} X_{2}^{A B}
\end{equation}
where
\begin{equation}
    \begin{split}
        X_{2}^{A B} = &\frac{1}{8} k \Big(4 \partial_{v}{K_{C E}}\bar{K}^{A B} \mu^{C E}-4 \partial_{v}{K_{C E}} \bar{K}\mu^{A B} \mu^{C E}-4 \partial_{v}{K_{C E}} \bar{K}^{B C}\mu^{A E}+\\
        & 4 \partial_{v}{K_{C E}} \bar{K}\mu^{A C} \mu^{B E}+4 \partial_{v}{K_{C E}} \bar{K}^{C E} \mu^{A B}+2D^{A}{K} \beta^{B}-4D^{C}{K} \beta_{C} \mu^{A B}-\\
        &2D^{C}{K^{A}\,_{C}} \beta^{B}+4D^{C}{K_{C}\,^{E}} \beta_{E} \mu^{A B}+4D^{C}{K^{A B}} \beta_{C}-2D^{A}{K^{B C}} \beta_{C}+2D^{A}{D^{B}{K}}-\\
        &4D^{C}{D^{E}{K_{C E}}} \mu^{A B}+2K^{A B} R[\mu]-4K^{A B} K^{C E} \bar{K}_{C E}-4K^{A C} R[\mu]^{B}\,_{C}-\\
        &2K^{C E} \beta_{C} \beta_{E} \mu^{A B}+K \beta^{C} \beta_{C} \mu^{A B}+K^{A C} \beta^{B} \beta_{C} -K^{A B} \beta^{C} \beta_{C}-4K^{C E} R[\mu]^{A}\,_{C}\,^{B}\,_{E}+\\
        &4K^{C E} R[\mu]_{C E} \mu^{A B}+4K^{C E} K_{C E} \bar{K} \mu^{A B}+8K^{A C} K^{B E} \bar{K}_{C E}-4K^{A C} K^{B}\,_{C} \bar{K}-\\
        &4K^{C E} K_{C}\,^{F} \bar{K}_{E F} \mu^{A B}-4\bar{K}^{A C} \partial_{v}{K_{C E}} \mu^{B E}+2D^{B}{K} \beta^{A}-2D^{C}{K^{B}\,_{C}} \beta^{A}-\\
        &2D^{B}{K^{A C}} \beta_{C}+2D^{B}{D^{A}{K}}-4K^{B C} R[\mu]^{A}\,_{C}+K^{B C} \beta^{A} \beta_{C}\Big),
    \end{split}
\end{equation}
\begin{equation}
    Y_{2}^{A} =k\left( D^{A}{K} K-2D_{B}{K} K^{A B}+D^{B}{K_{B C}} K^{A C}-D^{A}{K_{B C}} K^{B C}+D_{B}{K^{A}\,_{C}} K^{B C} \right)
\end{equation}
As expected from Section \ref{Classify}, we find $\varsigma_2^v=0$, i.e., the IWWHKR entropy coincides with the IWW entropy to $O(l^2)$ \cite{Hollands:2022}.

In completing the square in (\ref{EGBl2}) we have produced $O(l^4)$ terms, namely $
- l^4 X_{2 A B} X_{2}^{A B}$. We denote these as $l^4 F_4$. We can expand these out and proceed with the HKR algorithm at order $l^4$. We find
\begin{equation}\label{F4EGB}
    F_4=\partial_{v}\left[\frac{1}{\sqrt{\mu}} \partial_{v}\left(\sqrt{\mu} \varsigma^{v}_{4}\right)\right] + 2 K_{A B} X_{4}^{A B} + D_{A}{Y_{4}^{A}} + O(l^2)
\end{equation}
where\footnote{
For $d=4$ the GB term is topological. In this case
 one can show that $\zeta_4^v$ vanishes, using special identities satisfied by $2\times 2$ matrices.
} 
\begin{equation}
    \begin{split}
        \varsigma^{v}_{4} =& \frac{1}{8} k^{2} \Big[\left(6-d\right) K K \bar{K} \bar{K}-K K \bar{K}^{A B} \bar{K}_{A B}+4K K^{A B} \bar{K}_{A}\,^{C} \bar{K}_{B C}+\\
        &\left(-14+2d\right) K K^{A B} \bar{K} \bar{K}_{A B}-2K^{A B} K_{A}\,^{C} \bar{K}_{B}\,^{E} \bar{K}_{C E}-2K^{A B} K^{C E} \bar{K}_{A C} \bar{K}_{B E}+\\
        &\left(6-d\right) K^{A B} K^{C E} \bar{K}_{A B} \bar{K}_{C E}+4K^{A B} K_{A}\,^{C} \bar{K} \bar{K}_{B C}-K^{A B} K_{A B} \bar{K} \bar{K}\Big]
    \end{split}
\end{equation}
and $X_{4}^{A B}$ and $Y^{A}_{4}$ are very lengthy expressions of boost weight $+1$ and $+2$ respectively. Non-allowed terms do appear at this order after extracting $\varsigma^{v}_{4}$, and swapping them out using the equations of motion produces the $O(l^2)$ terms in (\ref{F4EGB}).

We then group like terms together and complete the square to write $F$ in the desired form:
\begin{multline}
    F = \partial_{v}\left[\frac{1}{\sqrt{\mu}} \partial_{v}\left(\sqrt{\mu}l^4 \varsigma^{v}_{4}\right)\right] \\
    +\left(K_{A B}+l^2 X_{2 A B} + l^4 X_{4 A B}\right) \left(K^{A B}+l^2 X_{2}^{A B} + l^6 X_{4}^{A B}\right) + D_{A}{ (l^2 Y_{2}^{A} + l^4 Y_{4}^{A}) } + O(l^6)
\end{multline}
The IWWHKR entropy density is then
\begin{equation}
    S^{v}= s^{v}_{IWW} + l^4 \varsigma^{v}_{4} + O(l^6)
\end{equation}
To this order, $S^{v}$ is gauge invariant as expected (section \ref{Classify}), as it only contains $\mu^{A B}, R[\mu], K_{A B}$ and $\bar{K}_{A B}$ which transform homogeneously under a change of GNCs on $C$.  

We can continue the algorithm to the next order by expanding out all the $O(l^6)$ terms. We find that the $O(l^6)$ part of $S^v$ is extremely lengthy. It is also not gauge invariant, however since it is so unwieldy we leave the discussion of gauge non-invariance to the considerably shorter expression arising from the quartic Lagrangian below.

\subsection{Cubic Order Riemann Lagrangians}\label{CubicCalc}

Let us now specialise to $d=4$. As mentioned above, the EGB term is purely topological in this dimension so we shall ignore it, and we can eliminate all other 4-derivative corrections through field redefinitions and total derivatives. At 6-derivative order, we can similarly reduce the number of corrections to just two \cite{Endlich:2017, Cano:2019}. The Lagrangian is $\mathcal{L}+O(l^6)$ where
\begin{equation}\label{4DEFT}
    \mathcal{L} = -2 \Lambda + R + l^4( k_{1} \mathcal{L}_{even} + k_{2} \mathcal{L}_{odd} ) 
\end{equation}
where $\mathcal{L}_{even}$ and $\mathcal{L}_{odd}$ are even and odd parity terms respectively, given by 
\begin{equation}
    \begin{split}
        \mathcal{L}_{even} =& R_{\mu \nu \kappa \lambda} R^{\kappa \lambda \chi \eta} R_{\chi \eta}\,^{\mu \nu} \\
        \mathcal{L}_{odd} =& R_{\mu \nu \kappa \lambda} R^{\kappa \lambda \chi \eta} R_{\chi \eta \rho \sigma} \epsilon^{\mu \nu \rho \sigma}
    \end{split}
\end{equation}
Let us follow our implementation of the HKR algorithm to find the entropy density at order $l^4$. The equation of motion for this Lagrangian is \eqref{eofm} with $N=6$ and 
\begin{equation}
    \begin{split}
        E_{\mu \nu} = &- \Lambda g_{\mu \nu} - G_{\mu \nu} +\\
        & l^4 \bigg[ k_{1} \Big( \frac{1}{2}R^{\kappa \lambda \alpha \beta} R_{\kappa \lambda}\,^{\rho \sigma} R_{\alpha \beta \rho \sigma} g_{\mu \nu}-3R_{\mu}\,^{\kappa \alpha \beta} R_{\nu \kappa}\,^{\rho \sigma} R_{\alpha \beta \rho \sigma} - 3R_{\mu}\,^{\alpha \beta \rho} \nabla_{\alpha}\nabla^{\sigma}{R_{\nu \sigma \beta \rho}}-\\
        & 6\nabla^{\alpha}{R_{\mu \alpha}\,^{\beta \rho}} \nabla^{\sigma}{R_{\nu \sigma \beta \rho}}-6\nabla^{\alpha}{R_{\mu}\,^{\beta \rho \sigma}} \nabla_{\beta}{R_{\nu \alpha \rho \sigma}} - 3R_{\mu}\,^{\alpha \beta \rho} \nabla^{\sigma}\nabla_{\alpha}{R_{\nu \sigma \beta \rho}}-\\
        & 3R_{\nu}\,^{\alpha \beta \rho} \nabla_{\alpha}\nabla^{\sigma}{R_{\mu \sigma \beta \rho}}-3R_{\nu}\,^{\alpha \beta \rho} \nabla^{\sigma}\nabla_{\alpha}{R_{\mu \sigma \beta \rho}} \Big) + \\
        & k_{2} \Big( - R_{\mu}\,^{\eta}\,_{\kappa \lambda} R_{\nu \eta}\,^{\alpha \beta} R_{\alpha \beta \rho \sigma} \epsilon^{\kappa \lambda \rho \sigma} - \frac{1}{2}R_{\mu \eta}\,^{\kappa \lambda} R_{\alpha \beta}\,^{\rho \sigma} R_{\kappa \lambda \rho \sigma} \epsilon^{\eta \alpha \beta}\,_{\nu} - R_{\mu}\,^{\kappa}\,_{\lambda \alpha} \epsilon^{\lambda \alpha \beta \rho} \nabla_{\kappa}\nabla^{\sigma}{R_{\nu \sigma \beta \rho}}-\\
        & 2\epsilon^{\kappa \lambda \alpha \beta} \nabla^{\rho}{R_{\mu \rho \kappa \lambda}} \nabla^{\sigma}{R_{\nu \sigma \alpha \beta}} - 2\epsilon^{\kappa \lambda \alpha \beta} \nabla^{\rho}{R_{\mu}\,^{\sigma}\,_{\kappa \lambda}} \nabla_{\sigma}{R_{\nu \rho \alpha \beta}} - R_{\mu}\,^{\kappa}\,_{\lambda \alpha} \epsilon^{\lambda \alpha \beta \rho} \nabla^{\sigma}\nabla_{\kappa}{R_{\nu \sigma \beta \rho}} + \\
        & R_{\kappa \lambda}\,^{\alpha \beta} \epsilon^{\kappa \lambda \rho}\,_{\nu} \nabla^{\sigma}\nabla_{\rho}{R_{\mu \sigma \alpha \beta}} + 2\epsilon^{\kappa \lambda \alpha}\,_{\nu} \nabla_{\kappa}{R_{\mu}\,^{\beta \rho \sigma}} \nabla_{\beta}{R_{\lambda \alpha \rho \sigma}} + 2\epsilon^{\kappa \lambda \alpha}\,_{\nu} \nabla^{\beta}{R_{\mu \beta}\,^{\rho \sigma}} \nabla_{\kappa}{R_{\lambda \alpha \rho \sigma}} + \\
        & R_{\mu}\,^{\kappa \lambda \alpha} \epsilon^{\beta \rho \sigma}\,_{\nu} \nabla_{\kappa}\nabla_{\beta}{R_{\lambda \alpha \rho \sigma}} + R_{\kappa \lambda}\,^{\alpha \beta} \epsilon^{\kappa \lambda \rho}\,_{\nu} \nabla_{\rho}\nabla^{\sigma}{R_{\mu \sigma \alpha \beta}} + R_{\mu}\,^{\kappa \lambda \alpha} \epsilon^{\beta \rho \sigma}\,_{\nu} \nabla_{\beta}\nabla_{\kappa}{R_{\lambda \alpha \rho \sigma}}-\\
        & R_{\mu}\,^{\eta \kappa \lambda} R_{\nu \eta \alpha \beta} R_{\kappa \lambda \rho \sigma} \epsilon^{\alpha \beta \rho \sigma} - \frac{1}{2}R_{\nu \eta}\,^{\kappa \lambda} R_{\alpha \beta}\,^{\rho \sigma} R_{\kappa \lambda \rho \sigma} \epsilon^{\eta \alpha \beta}\,_{\mu} - R_{\nu}\,^{\kappa}\,_{\lambda \alpha} \epsilon^{\lambda \alpha \beta \rho} \nabla_{\kappa}\nabla^{\sigma}{R_{\mu \sigma \beta \rho}} - \\
        & R_{\nu}\,^{\kappa}\,_{\lambda \alpha} \epsilon^{\lambda \alpha \beta \rho} \nabla^{\sigma}\nabla_{\kappa}{R_{\mu \sigma \beta \rho}} + R_{\kappa \lambda}\,^{\alpha \beta} \epsilon^{\kappa \lambda \rho}\,_{\mu} \nabla^{\sigma}\nabla_{\rho}{R_{\nu \sigma \alpha \beta}} + 2\epsilon^{\kappa \lambda \alpha}\,_{\mu} \nabla_{\kappa}{R_{\nu}\,^{\beta \rho \sigma}} \nabla_{\beta}{R_{\lambda \alpha \rho \sigma}} + \\
        & 2\epsilon^{\kappa \lambda \alpha}\,_{\mu} \nabla^{\beta}{R_{\nu \beta}\,^{\rho \sigma}} \nabla_{\kappa}{R_{\lambda \alpha \rho \sigma}} + R_{\nu}\,^{\kappa \lambda \alpha} \epsilon^{\beta \rho \sigma}\,_{\mu} \nabla_{\kappa}\nabla_{\beta}{R_{\lambda \alpha \rho \sigma}} + R_{\kappa \lambda}\,^{\alpha \beta} \epsilon^{\kappa \lambda \rho}\,_{\mu} \nabla_{\rho}\nabla^{\sigma}{R_{\nu \sigma \alpha \beta}} + \\
        & R_{\nu}\,^{\kappa \lambda \alpha} \epsilon^{\beta \rho \sigma}\,_{\mu} \nabla_{\beta}\nabla_{\kappa}{R_{\lambda \alpha \rho \sigma}} \Big) \bigg]
    \end{split}
\end{equation}
We now must find the IWW entropy current for this theory. $\mathcal{L}$ depends only on the Riemann tensor and not its derivatives up to and including order $l^4$, and hence we can calculate the IWW entropy density $s^{v}_{IWW}$ using the formula (\ref{WallIWWEFT}). Splitting this into the individual contributions from $\mathcal{L}_{even}$ and $\mathcal{L}_{odd}$ we get
\begin{equation}
    s^{v}_{IWW} = 1 + l^4( k_{1} s^{v}_{even} + k_{2} s^{v}_{odd} ) - s^v_{quadratic} 
\end{equation}
where
\begin{equation}
\label{evenodd}
    \begin{split}
    s^{v}_{even} =&-6R_{r v A B} R_{r v C D} \mu^{A C} \mu^{B D}-24R_{r v r A} R_{r v v B} \mu^{A B}+12R_{r v r v} R_{r v r v}-24 R_{r A v B} K^{A C} \bar{K}^{B}\,_{C} \\
    s^{v}_{odd} =& -4R_{A B C E} R_{r v F G} \epsilon^{A B} \mu^{C F} \mu^{E G}-8R_{r A B C} R_{r v v D} \epsilon^{B C} \mu^{A D}-8R_{r v r A} R_{v B C D} \epsilon^{C D} \mu^{A B}-\\
    &16R_{r v r A} R_{r v v B} \epsilon^{A B}+16R_{r v A B} R_{r v r v} \epsilon^{A B}+16 R_{r A v B} K^{A C} \bar{K}_{C D} \epsilon^{B D} +\\
    & 16 R_{r A v B} K^{A}\,_{C} \bar{K}^{B}\,_{D} \epsilon^{C D}-16 R_{r A v B} K_{C D} \bar{K}^{B D} \epsilon^{A C}
    \end{split}
\end{equation}
and where $s^{v}_{quadratic}$ is there to cancel any terms that are quadratic or higher in positive boost weight quantities when the Riemann components above are expanded out in GNCs, so that $s^v_{IWW}$ is only linear or zero order in positive boost weight. After expanding, we find
\begin{equation}
\begin{split}
    s^{v}_{quadratic} =& l^4\Big[ k_1 (12K^{A B} K^{C D} \bar{K}_{A C} \bar{K}_{B D}-36K^{A B} K_{A}\,^{C} \bar{K}_{B}\,^{D} \bar{K}_{C D}) + \\
    &k_2( 32K_{A}\,^{B} K_{B}\,^{C} \bar{K}_{E}\,^{F} \bar{K}_{C F} \epsilon^{A E}-16K_{A}\,^{B} K^{C E} \bar{K}_{B E} \bar{K}_{C F} \epsilon^{A F} ) \Big]
\end{split}
\end{equation}
The quantities $s^v_{even}$, $s^v_{odd}$ and $s^v_{quadratic}$ are separately gauge invariant. This is because each of them is a zero boost weight quantity depending only on Riemann components, $\mu^{A B}$, $\epsilon^{A B}$, $K_{A B}$ and $\bar{K}_{A B}$, all of which transform homogeneously on $C$. Hence $s^v_{IWW}$ is also gauge invariant without the need to add terms of quadratic or higher order in positive boost weight quantities (discussed at the end of section \ref{sec:IWW}). The gauge invariance of $s^v_{even}$ and $s^v_{odd}$ follows from the gauge invariance of the first part of \eqref{WallIWW}. What is perhaps surprising here is that $s^v_{quadratic}$ is also gauge invariant. We shall discuss this further in section \ref{GaugeInvariance}.

We can expand out the Riemann components in GNCs on $\mathcal{N}$ using the expressions in the Appendix of \cite{Hollands:2022} to get
\begin{equation}
    \begin{split}
        s^{v}_{IWW} =& 1 + \\
        &l^4 \bigg[ k_{1} \Big( 12{\alpha}^{2} -3D^{A}{\beta^{B}} D_{A}{\beta_{B}}+3D^{A}{\beta^{B}} D_{B}{\beta_{A}}+ 6\alpha \beta^{A} \beta_{A}+ \frac{3}{4}\beta^{A} \beta_{A} \beta^{B} \beta_{B} + \\
        &24D^{A}{\beta^{B}} K_{B}\,^{C} \bar{K}_{A C} - 12D^{A}{\beta^{C}} K_{A}\,^{B} \bar{K}_{C B}-12\mu^{A B} \partial_{r}{\beta_{A}} \partial_{v}{\beta_{B}}-\\
        &12K^{A B} \beta_{A} \partial_{r}{\beta_{B}}+ 6\bar{K}^{A B} \beta_{A} \partial_{v}{\beta_{B}}+ 12K^{A B} \bar{K}_{A}\,^{C} \beta_{B} \beta_{C} +24K^{A B} \bar{K}_{A}\,^{C} \partial_{v}{\bar{K}_{B C}}\Big)\\
        & k_{2} \Big( -4D_{A}{\beta_{B}} R[\mu] \epsilon^{A B} + 16D_{A}{\beta_{B}} \alpha \epsilon^{A B}+4D_{A}{\beta_{B}} \beta^{C} \beta_{C} \epsilon^{A B} +\\
        &8D^{A}{\beta^{B}} K_{A C} \bar{K}_{B E} \epsilon^{C E}+8D_{A}{\bar{K}_{B}\,^{C}} \epsilon^{A B} \partial_{v}{\beta_{C}}+8D_{A}{\bar{K}_{B}\,^{C}} K_{C}\,^{E} \beta_{E} \epsilon^{A B}-\\
        &4\bar{K}_{A}\,^{B} \beta_{C} \epsilon^{A C} \partial_{v}{\beta_{B}}+16D_{A}{K_{B}\,^{C}} \epsilon^{A B} \partial_{r}{\beta_{C}}+8K_{A}\,^{B} \beta_{C} \epsilon^{A C} \partial_{r}{\beta_{B}}-\\
        &8D_{A}{K_{B}\,^{C}} \bar{K}_{C}\,^{E} \beta_{E} \epsilon^{A B}-8K_{A}\,^{B} \bar{K}_{B}\,^{C} \beta_{C} \beta_{E} \epsilon^{A E}-8\epsilon^{A B} \partial_{r}{\beta_{A}} \partial_{v}{\beta_{B}}+\\
        &8K_{A}\,^{B} \beta_{B} \epsilon^{A C} \partial_{r}{\beta_{C}}+4\bar{K}_{A}\,^{B} \beta_{B} \epsilon^{A C} \partial_{v}{\beta_{C}}-8K_{A}\,^{B} \bar{K}_{C}\,^{E} \beta_{B} \beta_{E} \epsilon^{A C}-\\
        &8K_{A}\,^{B} \bar{K}_{B C} R[\mu] \epsilon^{A C}-16D^{A}{\beta^{B}} K_{B C} \bar{K}_{A E} \epsilon^{C E} +32K_{A}\,^{B} \bar{K}_{B C} \alpha \epsilon^{A C}+\\
        &8K_{A}\,^{B} \bar{K}_{B C} \beta^{E} \beta_{E} \epsilon^{A C}-8D_{A}{\beta^{B}} K_{B}\,^{C} \bar{K}_{C E} \epsilon^{A E}+16K^{A B} \bar{K}_{A C} \epsilon^{C E} \partial_{v}{\bar{K}_{B E}}-\\
        &16K_{A}\,^{B} \bar{K}_{C}\,^{E} \epsilon^{A C} \partial_{v}{\bar{K}_{B E}}+8D^{A}{\beta_{B}} K_{C}\,^{E} \bar{K}_{A E} \epsilon^{B C}-16K_{A}\,^{B} \bar{K}_{B}\,^{C} \epsilon^{A E} \partial_{v}{\bar{K}_{C E}} \Big) \bigg] 
    \end{split}
\end{equation}
In this expression, the terms involving only boost-weight $0$ quantities give the Iyer-Wald entropy. The other terms, which are linear in positive boost weight quantities, are the Wall terms. We can calculate $s^A$ using the method in \cite{Bhattacharyya:2021jhr}. The expression is very lengthy and is given in Appendix \ref{app:sAcubic}.

Proceeding with the HKR algorithm, we calculate $F \equiv E_{v v} - \partial_{v}\left[\frac{1}{\sqrt{\mu}} \partial_{v}\left(\sqrt{\mu} s^{v}_{IWW}\right) + D_{A}{ s^A }\right] $ in GNCs on $\mathcal{N}$, swap out any non-allowed terms using the equations of motion and then manipulate the order $l^4$ terms into the required form:
\begin{equation} \label{finalR3}
    F=\partial_{v}\left[\frac{1}{\sqrt{\mu}} \partial_{v}\left(\sqrt{\mu} l^4 \varsigma^{v}_{4}\right)\right] +\left(K_{A B}+l^4 X_{4 A B}\right) \left(K^{A B}+l^4 X_{4}^{A B}\right) + D_{A}(l^4 Y_{4}^{A}) + O(l^6)
\end{equation}
where we find
\begin{equation}
    \begin{split}
   \varsigma^{v}_4 =& k_{1} \Big(-36K^{A B} K_{A}\,^{C} \bar{K}_{B}\,^{E} \bar{K}_{C E}+6K K^{A B} \bar{K}_{A}\,^{C} \bar{K}_{B C}+6K^{A B} K_{A}\,^{C} \bar{K} \bar{K}_{B C} \Big) + \\
   & k_{2} \Big( 32K_{A}\,^{B} K_{B}\,^{C} \bar{K}_{E}\,^{F} \bar{K}_{C F} \epsilon^{A E}-8K K_{A}\,^{B} \bar{K}_{C}\,^{E} \bar{K}_{B E} \epsilon^{A C}+8K_{A}\,^{B} K^{C E} \bar{K}_{B F} \bar{K}_{C E} \epsilon^{A F}-\\
   &8K_{A}\,^{B} K_{B}\,^{C} \bar{K} \bar{K}_{C E} \epsilon^{A E} \Big)
    \end{split}
\end{equation}
and $X_{4}^{A B}$ and $Y_{4}^{A}$ are very lengthy expressions. The IWWHKR entropy density is then
\begin{equation}
    S^{v} = s^{v}_{IWW} + l^4 \varsigma^{v}_4 
\end{equation}
As expected (section \ref{Classify}), $\varsigma^{v}_4$ is gauge invariant since it involves only $\mu^{A B}$, $\epsilon^{A B}$, $K_{A B}$ and $\bar{K}_{A B}$. Thus, the leading order ($6$-derivative) EFT corrections to vacuum gravity in four dimensions produce a $S^{v}$ that is gauge invariant on $C$. It can be written in a manifestly gauge invariant way by re-instating the Riemann components in $s^v_{IWW}$: 
\begin{equation}
    \begin{split}
        S^{v} =& 1 + l^4 \Big[k_1 \Big(-6R_{r v A B} R_{r v C D} \mu^{A C} \mu^{B D}-24R_{r v r A} R_{r v v B} \mu^{A B}+12R_{r v r v} R_{r v r v}-\\
        &24 R_{r A v B} K^{A C} \bar{K}^{B}\,_{C}- 12K^{A B} K^{C D} \bar{K}_{A C} \bar{K}_{B D}+6K K^{A B} \bar{K}_{A}\,^{C} \bar{K}_{B C}+6K^{A B} K_{A}\,^{C} \bar{K} \bar{K}_{B C} \Big)\\
        &k_2 \Big( -4R_{A B C E} R_{r v F G} \epsilon^{A B} \mu^{C F} \mu^{E G}-8R_{r A B C} R_{r v v D} \epsilon^{B C} \mu^{A D}-8R_{r v r A} R_{v B C D} \epsilon^{C D} \mu^{A B}-\\
        &16R_{r v r A} R_{r v v B} \epsilon^{A B}+16R_{r v A B} R_{r v r v} \epsilon^{A B}+16 R_{r A v B} K^{A C} \bar{K}_{C D} \epsilon^{B D} +\\
        & 16 R_{r A v B} K^{A}\,_{C} \bar{K}^{B}\,_{D} \epsilon^{C D}-16 R_{r A v B} K_{C D} \bar{K}^{B D} \epsilon^{A C} -8 K K_{A}\,^{B} \bar{K}_{C}\,^{E} \bar{K}_{B E} \epsilon^{A C}+\\
        &8K_{A}\,^{B} K^{C E} \bar{K}_{B F} \bar{K}_{C E} \epsilon^{A F}-8K_{A}\,^{B} K_{B}\,^{C} \bar{K}_{C E} \bar{K} \epsilon^{A E}+16K_{A}\,^{B} K^{C E} \bar{K}_{B E} \bar{K}_{C F} \epsilon^{A F} \Big) \Big] 
    \end{split}
\end{equation}
This satisfies the second law, to quadratic order, modulo terms of order $l^6$.

We have followed a ``strict'' definition of the IWW entropy density, in which it contains terms only of up to linear order in positive boost weight quantities. An alternative definition might allow terms of quadratic or higher order in such quantities (which do not affect the linearized second law). For example, Wall defines the entropy for the above theory to be $s^v_{IWW} + s^v_{quadratic}$, corresponding to the just the first part of \eqref{WallIWW}. Adopting this alternative definition for the IWW entropy would not affect our final answer for $S^v$ because adding a quantity to $s^v_{IWW}$ and then running the HKR algorithm simply results in subtracting the same quantity from $\sum_n l^n \varsigma_n^v$ and so this quantity cancels out in the definition of $S^v$, equation \eqref{IWWHKRdef}.

\subsection{An Example Quartic Riemann Theory}\label{QuarticCalc}

We can go beyond 6-derivative, i.e. $O(l^4)$, terms in our EFT Lagrangian in $d=4$ and consider the next order, 8-derivative terms, i.e. $O(l^6)$. To do this in the most general case, we should include all possible 8-derivative terms in the Lagrangian, up to field redefinitions and topological terms. We should also be aware that the algorithm used to construct $\varsigma^{v}_{4}$ above produces higher order terms via two sources: (a) swapping non-allowed terms using the higher order equations of motion and (b) the remainder, $-l^8 X_{4 A B} X_{4}^{A B}$, from completing the square in (\ref{finalR3}). However since there are no $O(l^2)$ terms in our theory, both of these sources produce terms of order at least $(l^4)^2 = l^8$. Therefore, the $O(l^6)$ terms in the entropy arise only from the 8-derivative terms in the Lagrangian. 

The minimal set of 8-derivative terms in EFT after field redefinitions and neglecting total derivatives is discussed in \cite{Endlich:2017}. Here we will restrict our attention to just one term since this is enough to demonstrate that the IWWHKR entropy can be gauge non-invariant. Therefore let us consider adding to our $d=4$ EFT Lagrangian (\ref{4DEFT}) the following 8-derivative term
\begin{equation}
    l^6 \mathcal{L}_8 = k l^6 R_{\mu \nu \rho \sigma} R^{\mu \nu \rho \sigma} R_{\kappa \lambda \chi \eta} R^{\kappa \lambda \chi \eta} 
\end{equation}
This contributes the following to the equation of motion:
\begin{equation}
    \begin{split}
    E^{(6)}_{\mu \nu} =& k l^6 \Big[ \frac{1}{2}R^{\chi \eta \kappa \lambda} R_{\chi \eta \kappa \lambda} R^{\alpha \beta \rho \sigma} R_{\alpha \beta \rho \sigma} g_{\mu \nu}-4R_{\mu}\,^{\eta \kappa \lambda} R_{\nu \eta \kappa \lambda} R^{\alpha \beta \rho \sigma} R_{\alpha \beta \rho \sigma}-\\
    & 4R^{\kappa \lambda \alpha \beta} R_{\kappa \lambda \alpha \beta} \nabla^{\rho}\left(\nabla^{\sigma}{R_{\mu \rho \nu \sigma}}\right)-16R^{\kappa \lambda \alpha \beta} \nabla^{\rho}{R_{\mu \rho \nu}\,^{\sigma}} \nabla_{\sigma}{R_{\kappa \lambda \alpha \beta}}-\\
    &8R_{\mu}\,^{\kappa}\,_{\nu}\,^{\lambda} R^{\alpha \beta \rho \sigma} \nabla_{\kappa}\left(\nabla_{\lambda}{R_{\alpha \beta \rho \sigma}}\right)-16R_{\mu}\,^{\kappa}\,_{\nu}\,^{\lambda} \nabla_{\kappa}{R^{\alpha \beta \rho \sigma}} \nabla_{\lambda}{R_{\alpha \beta \rho \sigma}}-\\
    &4R^{\kappa \lambda \alpha \beta} R_{\kappa \lambda \alpha \beta} \nabla^{\rho}\left(\nabla^{\sigma}{R_{\mu \sigma \nu \rho}}\right)-16R^{\kappa \lambda \alpha \beta} \nabla^{\rho}{R_{\mu}\,^{\sigma}\,_{\nu \rho}} \nabla_{\sigma}{R_{\kappa \lambda \alpha \beta}}-\\
    &8R_{\mu}\,^{\kappa}\,_{\nu}\,^{\lambda} R^{\alpha \beta \rho \sigma} \nabla_{\lambda}\left(\nabla_{\kappa}{R_{\alpha \beta \rho \sigma}}\right) \Big]
    \end{split}
\end{equation}
We can calculate the contribution to the IWW entropy current:
\begin{equation} \label{IWWR4}
    E^{(6)}_{v v} = \partial_{v}\left[\frac{1}{\sqrt{\mu}} \partial_{v}\left(\sqrt{\mu} s^{(6)v}_{IWW}\right) + D_{A}{ s^{(6)A} }\right] + ...
\end{equation}
where the ellipsis denotes terms at least quadratic in positive boost weight quantities. The lengthy expression for $s^{(6)A}$ is given in Appendix \ref{app:s6A}. For $s^{(6)v}_{IWW}$ we find
\begin{equation}\label{sIWW6}
    \begin{split}
        s^{(6)v}_{IWW}=&k l^6\Big(128R_{r v r A} R_{r v r v} R_{r v v B} \mu^{A B} -8R_{A B C E} R_{F G H I} R_{r v r v} \mu^{A F} \mu^{B G} \mu^{C H} \mu^{E I}-\\
        &64R_{r A B C} R_{r v r v} R_{v E F G} \mu^{A E} \mu^{B F} \mu^{C G}-64R_{r A v B} R_{r C v E} R_{r v r v} \mu^{A E} \mu^{B C}-\\
        &64R_{r A r B} R_{r v r v} R_{v C v E} \mu^{A C} \mu^{B E}+32R_{r v A B} R_{r v C E} R_{r v r v} \mu^{A C} \mu^{B E}-\\
        &32R_{r v r v} R_{r v r v} R_{r v r v}-8R_{A B C E} R_{F G H I} K^{J P} \bar{K}_{J P} \mu^{A F} \mu^{B G} \mu^{C H} \mu^{E I}-\\
        &64R_{r A B C} R_{v E F G} K^{H I} \bar{K}_{H I} \mu^{A E} \mu^{B F} \mu^{C G}-64R_{r A v B} R_{r C v E} K^{F G} \bar{K}_{F G} \mu^{A E} \mu^{B C}-\\
        &64R_{r A r B} R_{v C v E} K^{F G} \bar{K}_{F G} \mu^{A C} \mu^{B E}-64R_{r A r B} R_{v C v E} K^{A B} \bar{K}^{C E}+\\
        &32R_{r v A B} R_{r v C E} K^{F G} \bar{K}_{F G} \mu^{A C} \mu^{B E}+128R_{r v r A} R_{r v v B} K^{C E} \bar{K}_{C E} \mu^{A B}-\\
        &32R_{r v r v} R_{r v r v} K^{A B} \bar{K}_{A B} \Big) - s^{(6)v}_{quadratic}
\end{split}
\end{equation}
where we cancel the terms that are of quadratic or higher order in positive boost weight quantities with
\begin{equation}
\label{s6quad}
\begin{split}
        s^{(6)v}_{quadratic} =& k l^6 \Big(-32K^{A B} K_{A B} \bar{K}^{C E} \bar{K}_{C E} \alpha-8K^{A B} K_{A B} \bar{K}^{C E} \bar{K}_{C E} \beta^{F} \beta_{F}-\\
        &32K^{A B} K^{C E} \bar{K}_{A B} \bar{K}_{C E} \alpha+64K^{A B} K_{A}\,^{C} \bar{K}_{B}\,^{E} \bar{K}_{C E} \alpha+\\
        &16K^{A B} K_{A}\,^{C} \bar{K}_{B}\,^{E} \bar{K}_{C E} \beta^{F} \beta_{F}-128K^{A B} K^{C E} \bar{K}_{A C} \bar{K}_{B E} \alpha+\\
        &64K^{A B} K_{A}\,^{C} \alpha \partial_{r}{\bar{K}_{B C}}+16K^{A B} K_{A}\,^{C} \beta^{E} \beta_{E} \partial_{r}{\bar{K}_{B C}}+32K K^{B C} \bar{K}_{B C} \bar{K} R[\mu]-\\
        &32K^{A B} K^{C E} \bar{K}_{A B} \bar{K}_{C E} R[\mu]-32K^{A B} K_{A B} K^{C E} \bar{K}_{C E} \bar{K}^{F G} \bar{K}_{F G}-\\
        &32K^{A B} K^{C E} K^{F G} \bar{K}_{A B} \bar{K}_{C E} \bar{K}_{F G}+64K^{A B} K_{A}\,^{C} K^{E F} \bar{K}_{B}\,^{G} \bar{K}_{C G} \bar{K}_{E F}-\\
        &128D^{A}{K^{B C}} D_{A}{\bar{K}_{B C}} K^{E F} \bar{K}_{E F}+128D^{A}{K^{B C}} D_{B}{\bar{K}_{A C}} K^{E F} \bar{K}_{E F}+\\
        &64D^{A}{\bar{K}^{B C}} K_{B C} K^{E F} \bar{K}_{E F} \beta_{A}-64D^{A}{\bar{K}^{B C}} K_{A B} K^{E F} \bar{K}_{E F} \beta_{C}-\\
        &64D^{A}{K^{B C}} K^{E F} \bar{K}_{B C} \bar{K}_{E F} \beta_{A}+64D^{A}{K^{B C}} K^{E F} \bar{K}_{A B} \bar{K}_{E F} \beta_{C}-\\
        &32K^{A B} K^{C E} \bar{K}_{A B} \bar{K}_{C}\,^{F} \beta_{E} \beta_{F}+128D^{A}{\beta^{B}} K_{A}\,^{C} K^{E F} \bar{K}_{B C} \bar{K}_{E F}+\\
        &128K^{A B} K^{C E} \bar{K}_{A B} \bar{K}_{C}\,^{F} \partial_{v}{\bar{K}_{E F}}-128K^{A B} K^{C E} K^{F G} \bar{K}_{A B} \bar{K}_{C F} \bar{K}_{E G}-\\
        &64K^{A B} \bar{K}_{A B} \mu^{C E} \mu^{F G} \partial_{r}{\bar{K}_{C F}} \partial_{v}{K_{E G}}+64K^{A B} K_{A}\,^{C} K^{E F} \bar{K}_{E F} \partial_{r}{\bar{K}_{B C}}+\\
        &64K^{A B} \bar{K}_{A B} \bar{K}^{C E} \bar{K}_{C}\,^{F} \partial_{v}{K_{E F}}-64K^{A B} \bar{K}^{C E} \partial_{r}{\bar{K}_{A B}} \partial_{v}{K_{C E}}+\\
        &64K^{A B} K_{A}\,^{C} K^{E F} \bar{K}_{B C} \partial_{r}{\bar{K}_{E F}}+64K^{A B} \bar{K}_{A}\,^{C} \bar{K}_{B C} \bar{K}^{E F} \partial_{v}{K_{E F}}-\\
        &64K^{A B} K_{A}\,^{C} K^{E F} \bar{K}_{B C} \bar{K}_{E}\,^{G} \bar{K}_{F G}-64D^{A}{\beta^{B}} K_{B}\,^{C} K^{E F} \bar{K}_{A C} \bar{K}_{E F}+\\
        &64K^{A B} \bar{K}_{A B} \mu^{C E} \partial_{r}{\beta_{C}} \partial_{v}{\beta_{E}}+64K^{A B} K^{C E} \bar{K}_{A B} \beta_{C} \partial_{r}{\beta_{E}}-\\
        &32K^{A B} \bar{K}_{A B} \bar{K}^{C E} \beta_{C} \partial_{v}{\beta_{E}} -32K^{A B} K^{C E} \bar{K}_{A C} \bar{K}_{B E} \beta^{F} \beta_{F}\\
        &+24K^{A B} K^{C E} \bar{K}_{A B} \bar{K}_{C E} \beta^{F} \beta_{F}\Big)
    \end{split}
\end{equation}
We shall prove that $s^{(6)v}_{IWW}$ is gauge invariant in the next section. Proceeding with the HKR algorithm once again with $l^6 F_6 = E^{(6)}_{v v} - \partial_{v}\left[\frac{1}{\sqrt{\mu}} \partial_{v}\left(\sqrt{\mu} s^{(6)v}_{IWW}\right) + D_{A}{ s^{(6)A} }\right]$, we find
\begin{equation} \label{finalR4}
    F_6 = \partial_{v}\left[\frac{1}{\sqrt{\mu}} \partial_{v}\left(\sqrt{\mu} \varsigma^{v}_{6}\right)\right] + 2 K_{A B} X_{6}^{A B} + D_{A} Y_{6}^{A} + O(l^2)
\end{equation}
where
\begin{equation}
\label{varsigma6}
    \begin{split}
   \varsigma^{v}_6 =& \frac{k}{3} \Big[ - 64 K^{A B} \bar{K}^{C E} \partial_{r}{\bar{K}_{A B}} \partial_{v}{K_{C E}}+64K^{A B} \bar{K}_{A}\,^{C} \bar{K}_{B C} \bar{K}^{E F} \partial_{v}{K_{E F}} -\\
   & 64K^{A B} \bar{K}_{A B} \mu^{C E} \mu^{F G} \partial_{r}{\bar{K}_{C F}} \partial_{v}{K_{E G}}+64K^{A B} \bar{K}_{A B} \bar{K}^{C E} \bar{K}^{F}\,_{E} \partial_{v}{K_{C F}}+\\
   &96 D^{A}{D_{A}{K}} K^{B C} \partial_{r}{\bar{K}_{B C}}-96 D^{A}{D_{A}{K}} K^{B C} \bar{K}_{B}\,^{E} \bar{K}_{C E}-96 D^{A}{D_{A}{K^{B C}}} K \partial_{r}{\bar{K}_{B C}}+\\
   &96 D^{A}{D_{A}{K^{B C}}} K \bar{K}_{B}\,^{E} \bar{K}_{C E}-96 D^{A}{K^{B C}} K^{E F} \bar{K}_{B C} \bar{K}_{E F} \beta_{A}+48 D^{A}{K} K^{B C} \bar{K}_{B}\,^{E} \bar{K}_{C E} \beta_{A}-\\
   &192 D^{A}{K^{B C}} D_{A}{\bar{K}_{B C}} K^{E F} \bar{K}_{E F}+192 D^{A}{K^{B C}} D_{B}{\bar{K}_{A C}} K^{E F} \bar{K}_{E F}-48 D^{A}{K} K^{B C} \beta_{A} \partial_{r}{\bar{K}_{B C}}+\\
   &96 D^{A}{K^{B C}} K^{E F} \bar{K}_{A C} \bar{K}_{E F} \beta_{B}-96 D^{A}{K_{A}\,^{B}} K^{C E} \bar{K}_{B F} \bar{K}_{C E} \beta^{F}+96 D^{A}{K} K^{B C} \bar{K}_{A E} \bar{K}_{B C} \beta^{E}-\\
   &48 D^{A}{K^{B C}} K \beta_{A} \partial_{r}{\bar{K}_{B C}}+48 D^{A}{K^{B C}} K \bar{K}_{B}\,^{E} \bar{K}_{C E} \beta_{A}-192 D^{A}{K_{A}\,^{B}} D^{C}{\bar{K}_{B C}} K^{E F} \bar{K}_{E F}+\\
   &192 D^{A}{K_{A}\,^{B}} D_{B}{\bar{K}} K^{C E} \bar{K}_{C E}+96 D^{A}{K_{A}\,^{B}} K^{C E} \bar{K} \bar{K}_{C E} \beta_{B}+192 D^{A}{K} D^{B}{\bar{K}_{A B}} K^{C E} \bar{K}_{C E}-\\
   &192 D^{A}{K} D_{A}{\bar{K}} K^{B C} \bar{K}_{B C}-96 D^{A}{K} K^{B C} \bar{K} \bar{K}_{B C} \beta_{A}+96 D^{A}{K_{A}\,^{B}} D_{B}\left(\partial_{r}{\bar{K}_{C E}}\right) K^{C E}-\\
   &192 D^{A}{K_{A}\,^{B}} D_{B}{\bar{K}_{C E}} K^{F C} \bar{K}_{F}\,^{E}+96 D^{A}{K} D_{A}\left(\partial_{r}{\bar{K}_{B C}}\right) K^{B C}-192 D^{A}{K} D_{A}{\bar{K}_{B C}} K^{E B} \bar{K}_{E}\,^{C}+\\
   &96 D_{A}{K^{B C}} D_{E}\left(\partial_{r}{\bar{K}_{B C}}\right) K^{A E}-192 D_{A}{K^{B C}} D_{E}{\bar{K}_{C F}} K^{A E} \bar{K}_{B}\,^{F}-48 D_{A}{K^{B C}} K^{A E} \beta_{E} \partial_{r}{\bar{K}_{B C}}-\\
   &48 D^{A}{K_{A}\,^{B}} K^{C E} \beta_{B} \partial_{r}{\bar{K}_{C E}}+48 D_{A}{K^{B C}} K^{A E} \bar{K}_{B}\,^{F} \bar{K}_{C F} \beta_{E}+48 D^{A}{K_{A}\,^{B}} K^{C E} \bar{K}_{C}\,^{F} \bar{K}_{E F} \beta_{B}-\\
   &96 D^{A}{K^{B C}} D_{A}\left(\partial_{r}{\bar{K}_{B C}}\right) K+192 D^{A}{K^{B C}} D_{A}{\bar{K}_{C E}} K \bar{K}_{B}\,^{E} - 16 K K^{A B} R[\mu] \partial_{r}{\bar{K}_{A B}} -\\
   &16 D_{A}{\beta_{B}} K^{A B} K^{C E} \partial_{r}{\bar{K}_{C E}}+16 K^{A B} K^{C E} \beta_{A} \beta_{B} \partial_{r}{\bar{K}_{C E}}+32 K K^{A B} \Lambda \partial_{r}{\bar{K}_{A B}}+\\
   &16 K K^{A B} \bar{K}_{A}\,^{C} \bar{K}_{B C} R[\mu]+16 D_{A}{\beta_{B}} K^{A B} K^{C E} \bar{K}_{C}\,^{F} \bar{K}_{E F} - 16 K^{A B} K^{C E} \bar{K}_{A}\,^{F} \bar{K}_{B F} \beta_{C} \beta_{E} -\\
   &32 K K^{A B} \bar{K}_{A}\,^{C} \bar{K}_{B C} \Lambda-48 D^{A}{\beta_{A}} K K^{B C} \partial_{r}{\bar{K}_{B C}}+48 D^{A}{\beta_{A}} K K^{B C} \bar{K}_{B}\,^{E} \bar{K}_{C E}+\\
   &216 K^{A B} K^{C E} \bar{K}_{A B} \bar{K}_{C E} R[\mu]-60 K K^{A B} \bar{K} \bar{K}_{A B} R[\mu]+48 D_{A}{\bar{K}_{B C}} K^{B C} K^{E F} \bar{K}_{E F} \beta^{A} -\\
   &64 D_{A}{\bar{K}_{B C}} K^{A B} K^{E F} \bar{K}_{E F} \beta^{C}+28 K^{A B} K^{C E} \bar{K}_{A B} \bar{K}_{C E} \beta^{F} \beta_{F}-48 K K^{A B} \bar{K}_{A B} \bar{K}_{C E} \beta^{C} \beta^{E} -\\
   &112 D^{A}{\beta_{B}} K^{B C} K^{E F} \bar{K}_{A C} \bar{K}_{E F}+112 D_{A}{\beta^{B}} K^{A C} K^{E F} \bar{K}_{B C} \bar{K}_{E F}+8 D_{A}{\beta_{B}} K^{A B} K^{C E} \bar{K} \bar{K}_{C E} -\\
   &304 K^{A B} K^{C E} \bar{K}_{A B} \bar{K}_{C E} \Lambda+104 K K^{A B} \bar{K} \bar{K}_{A B} \Lambda-96 D^{A}{\bar{K}_{A B}} K K^{C E} \bar{K}_{C E} \beta^{B}+\\
   &96 D_{A}{\bar{K}} K K^{B C} \bar{K}_{B C} \beta^{A}+48 K K^{A B} \bar{K} \bar{K}_{A B} \beta^{C} \beta_{C}+64 D^{A}{\bar{K}_{A B}} K^{C B} K^{E F} \bar{K}_{E F} \beta_{C}-\\
   &48 D_{A}{\bar{K}} K^{B A} K^{C E} \bar{K}_{C E} \beta_{B} - 28 K^{A B} K^{C E} \bar{K} \bar{K}_{A B} \beta_{C} \beta_{E} - 16 D^{A}{D_{A}{\bar{K}_{B C}}} K^{B C} K^{E F} \bar{K}_{E F}+\\
   &16 D^{A}{D_{B}{\bar{K}_{A C}}} K^{B C} K^{E F} \bar{K}_{E F} - 8 D^{A}{\beta_{A}} K^{B C} K^{E F} \bar{K}_{B C} \bar{K}_{E F} - 16 D_{A}{D^{B}{\bar{K}_{B C}}} K^{A C} K^{E F} \bar{K}_{E F}+\\
   &16 D_{A}{D_{B}{\bar{K}}} K^{A B} K^{C E} \bar{K}_{C E}-48 D_{A}\left(\partial_{r}{\bar{K}_{B C}}\right) K K^{B C} \beta^{A}+96 D_{A}{\bar{K}_{B C}} K K^{E B} \bar{K}_{E}\,^{C} \beta^{A}-\\
   &96 K^{A B} K_{A}\,^{C} R[\mu] \partial_{r}{\bar{K}_{B C}}+96 D_{A}{D_{B}\left(\partial_{r}{\bar{K}_{C E}}\right)} K^{A B} K^{C E}-192 D_{A}{D_{B}{\bar{K}_{C E}}} K^{A B} K^{F C} \bar{K}_{F}\,^{E}-\\
   &192 D_{A}{\bar{K}_{B}\,^{C}} D_{E}{\bar{K}_{C F}} K^{A E} K^{B F}+192 K^{A B} K^{C E} \bar{K}_{A C} \bar{K}_{B E} R[\mu]+48 K^{A B} K_{A B} \bar{K}^{C E} \bar{K}_{C E} R[\mu]-\\
   &96 K^{A B} K_{A}\,^{C} \bar{K}_{B}\,^{E} \bar{K}_{C E} R[\mu]-48 D_{A}\left(\partial_{r}{\bar{K}_{B C}}\right) K^{B C} K^{E A} \beta_{E}+96 D_{A}{\bar{K}_{B C}} K^{E A} K^{F B} \bar{K}_{F}\,^{C} \beta_{E} \Big]
    \end{split}
\end{equation}
and $X_{6}^{A B}$ and $Y_{6}^{A}$ are (also) very lengthy expressions. The IWWHKR entropy for this 8-derivative theory is given by adding $s^{(6)v}_{IWW}+l^6 \varsigma^v_6$ to the result for the general 6-derivative theory in the previous section. The main reason for performing this calculation is to investigate whether or not the result is gauge invariant. This will be done in the next section. 

\section{Gauge (Non-)Invariance} \label{GaugeInvariance}

\subsection{Introduction}

In order to define our dynamical black hole entropy, we took a spacelike cross-section $C$ of the horizon $\mathcal{N}$ and chose Gaussian Null Co-ordinates $(r,v,x^A)$ with $r=v=0$ on $C$. This choice defines a foliation $C(v)$ of $\mathcal{N}$. The entropy of $C(v)$ is then given by the integral
\begin{equation}
    S_{IWWHKR}(v) = 4 \pi \int_{C(v)} \text{d}^{d-2}x \sqrt{\mu} S^{v} 
\end{equation}

If we fix $C$ then there are two freedoms in our choice of GNCs: (a) picking a different set of co-ordinates $x'^A$ on $C$, and (b) rescaling the affine parameter on each generator $v'=v/a(x^A)$ with $a(x^A)>0$. The procedure for calculating $S^v$ is manifestly covariant in $A, B, C,...$ indices and so (a) will not change $S_{IWWHKR}(v)$. If instead we make the rescaling (b) then in general our foliation will change $C'(v')\neq C(v)$. Therefore $S_{IWWHKR}(v)$ and $S'_{IWWHKR}(v')$ calculate the entropy of different surfaces, so we do not expect them to be the same. However, we should hope that $S_{IWWHKR}(v)$ is gauge invariant at $v=v'=0$, since $C(0)$ and $C'(0)$ are the same surface $C$. 

Hence we concern ourselves with how quantities change under the rescaling (b) when evaluated on $C$. The transformation laws of tensorial components and all allowed GNC quantities on $C$ are given in Section \ref{GaugeInvarianceLaws}. Most terms transform homogeneously: $T^{\mu_1 ... \mu_n}_{\nu_1 ... \nu_m}, \mu_{A B}, \epsilon_{A B}, R_{A B C D}[\mu], D_A, \partial_{v}^p K_{A B}$ and $\partial_{r}^p \Bar{K}_{A B}$ just gain a factor of $a^b$, where $b$ is their boost weight. However, $\beta_A$ transforms as
\begin{equation}
    \beta'_{A}= \beta_A+2D_{A} \log a
\end{equation}
and so the presence of $\beta_A$ is a warning sign of gauge non-invariance. Also note that a quantity such as $D_{A}{ K_{B C}}$ transforms as
\begin{equation}
\begin{split}
    D'_{A}{ K'_{B C} } =& D_{A}(a K_{B C})\\
    =& a (D_A K_{B C} + D_{A}( \log a ) K_{B C})
\end{split}
\end{equation}
which is also inhomogeneous. This will be the case for all $D$ derivatives of $\partial_{v}^p K_{A B}$ or $\partial_{r}^p \Bar{K}_{A B}$. We can get round this by swapping $D_A$ derivatives for gauge covariant derivatives $\mathcal{D}_A$ defined by \cite{Hollands:2022}
\begin{equation}
    \mathcal{D}_A T = D_A T -(b/2) \beta_A T
\end{equation}
for $T=\mathcal{D}_{A_1}...\mathcal{D}_{A_n} \partial_{v}^p K_{A B}$ or $\mathcal{D}_{A_1}...\mathcal{D}_{A_n} \partial_{r}^p \Bar{K}_{A B}$ with boost weight $b$. This can be shown to transform homogeneously as $\mathcal{D}'_A T'= a^b \mathcal{D}_A T$ on $C$. 

\subsection{IWW entropy}
We shall start by discussing how the IWW entropy behaves under a gauge transformation. For EGB theory, the IWW entropy is the same as the Iyer-Wald entropy, which is manifestly gauge invariant. For the cubic and quartic Riemann Lagrangians, the IWW entropy is determined by equation \eqref{WallIWW}. The first part of this equation is manifestly gauge invariant. Hence $s^v_{IWW} + s^v_{quadratic}$ must be gauge invariant, as confirmed by equations (\ref{evenodd}) and (\ref{sIWW6}). For the cubic theories we found that $s^v_{quadratic}$ is also gauge invariant, and hence so is $s^v_{IWW}$. The same is true for the quartic theory considered above. This can be shown as follows. In $s^{(6)v}_{quadratic}$ (equation \eqref{s6quad}) we swap out non-allowed terms for allowed terms plus Ricci components using the elimination rules in Appendix A. Here we do not eliminate the Ricci components using equations of motion, we are simply working with Ricci components since, as tensor components, they transform homogeneously under gauge transformations. We also swap $D$ derivatives for $\mathcal{D}$ derivatives on $K_{A B}$ and $\Bar{K}_{A B}$ terms. Of the terms that are left, the only ones that can transform inhomogeneously under a gauge transformation are those involving $D_{A_1}...D_{A_2} \beta_A$. We find that
\begin{equation}
    s^{(6)v}_{quadratic} = 192 k l^6K^{A B} K^{C E} \bar{K}_{A B} \Bar{K}_{C}\,^{F} D_{[E}{\beta_{F]}}+ \text{terms independent of $\beta_A$}
\end{equation}
But $D_{[A} \beta_{B]}$ is clearly gauge invariant (it is the ``field strength'' of the connection $\beta_A$). Hence $s^{(6)v}_{quadratic}$ is gauge invariant. In summary, for both the cubic and quartic Lagrangians, we have found that $s^v_{quadratic}$, and hence also $s^v_{IWW}$, is gauge invariant. 

This is puzzling. HKR proved that, without modification, the IWW entropy is gauge invariant to linear order in perturbation theory. Since the IWW entropy is of at most linear order in positive boost weight quantities, one might think that this automatically implies that is gauge invariant to all orders. That this is not true can be seen as follows. The easiest way to perform a gauge transformation of an expression involving (derivatives of) $\alpha,\beta_A$ etc is to rewrite it in terms of a new set of quantities (e.g. Ricci components) that transform nicely under gauge transformations. In general, this rewriting does not preserve the property of being ``linear in positive boost weight quantities.'' (For example the linear term $D_A \partial_v \beta_B$ can be eliminated in favour of $\nabla_A R_{vB}$ but this introduces nonlinear terms of the schematic form $K^2 \bar{K}$.) Hence after applying a gauge transformation, when we transform back to the original set of quantities, the difference $s^{v'}_{IWW}-s^v_{IWW}$ will involve not just terms of linear order in positive boost weight quantities, but also possibly terms of higher order. The former must vanish by the linear argument of \cite{Hollands:2022} but the latter may not. This problem is what the ``improvement'' terms of \cite{Hollands:2022} are designed to fix. But surprisingly, in the cases we have studied, such terms are not required, and $s^v_{IWW}$ is gauge invariant without improvement. 

To solve this puzzle, we shall now show that this result holds for any Lagrangian that depends only on the Riemann tensor (and not its derivatives). For such a Lagrangian, $s^v_{IWW}$ is given by \eqref{WallIWW}, which involves the Riemann tensor but not its derivatives. Since the Riemann tensor has dimension $2$, this implies that $s^v_{IWW}$ depends only on quantities with dimension of $2$ or less. Any such quantity is built from  ``primitive factors'' (in the terminology of \cite{Hollands:2022}) belonging to one of the following sets, where the subscript refers to the boost weight: $S_2=\{\partial_v K_{AB} \}$, $S_{-2} = \{ \partial_r \bar{K}_{AB} \}$, $S_1 = \{D_A K_{BC}, K_{AB}, \partial_v \beta_A\}$, $S_{-1}=\{ D_A \bar{K}_{BC}, \bar{K}_{AB}, \partial_r \beta_A \}$ and 
\begin{equation}
S_0=\{ \mu_{AB},\mu^{AB},\epsilon_{AB},\alpha,\beta_A, D_A \beta_B,R[\mu]_{ABCD}, \partial_v \partial_r \mu_{AB} \}.
\end{equation}
Furthermore, $s^v_{IWW}$ depends at most linearly on elements of $S_2$ and $S_1$. We can write $s^v_{IWW} =s^{v}_0 + s^{v}_1$ where $s^{v}_0$ is built only from terms of zero boost weight and $s^{v}_1$ is a sum of terms, each containing exactly one primitive factor of positive boost weight. Now, $s^{v}_{0}$ is simply the Iyer-Wald entropy density, which is gauge-invariant by definition. So we just need to understand how $s^{v}_{1}$ transforms. 

Using a formula from appendix \ref{app:elimination}, $\partial_v \beta_A$ can be eliminated in favour of $R_{vA}$ and other elements of the above sets. Importantly, $\partial_v \beta_A$ depends linearly on $R_{vA}$ and other quantities with positive boost weight so when we eliminate it, we do not introduce any nonlinear dependence on positive boost weight quantities (unlike what happens for a term like $D_A \partial_v \beta_B$, mentioned above, which might arise from a more general Lagrangian involving derivatives of the Riemann tensor). Similarly we can eliminate $\partial_r \beta_A$ in favour of $R_{rA}$ and other quantities listed above. We can also eliminate $D_A K_{BC}$ and $D_A \bar{K}_{BC}$ in favour of $\mathcal{D}_A K_{BC}$ and $\mathcal{D}_A \bar{K}_{BC}$ respectively. The result is to replace $S_1$ and $S_{-1}$ with $S'_1 = \{\mathcal{D}_A K_{BC}, K_{AB}, R_{vA}\}$ and $S_{-1}'=\{ \mathcal{D}_A \bar{K}_{BC}, \bar{K}_{AB}, R_{rA} \}$ respectively. So now $s^v_1$ is a sum of terms built from quantities belonging to $S_{\pm 2}$, $S_{\pm 1}'$ and $S_0$, and each term in $s^v_1$ contains exactly one element of $S_2$ or $S_1'$. We can write 
 \begin{equation}
 s^{v}_{1}=\sum_i P_i N_i Z_i
 \end{equation}
 where $P_i$ is an element of $S_2$ or $S_1'$, with boost weight $b_i \in \{2,1\}$, $N_i$ has boost weight $-b_i$ and is either an element of $S_{-2}$ or $S_{-1}'$ or a product of two elements in $S_{-1}'$. $Z_i$ has boost weight $0$ and is a product of elements of $S_0$. Now $P_i$ and $N_i$ are quantities that transform homogeneously under a gauge transformation and $P_i N_i$ has boost weight zero so it is gauge invariant. Hence, under a gauge transformation the change in $s^v_1$ is
 \begin{equation}
 \label{Deltas}
 \Delta s^{v}_{1}=\sum_i P_i N_i \Delta Z_i
 \end{equation}
 where $\Delta Z_i$ is the change in $Z_i$, which arises from the dependence of $Z_i$ on the quantities $\alpha,\beta_A,D_A \beta_B$ and $\partial_v \partial_r \mu_{AB}$ which do not transform homogeneously under a gauge transformation. Importantly, the transformation laws for these quantities involve only other quantities of boost weight $0$ and not, say, quantities like $K \bar{K}$. This is obvious for $\beta_A$ and $D_A \beta_B$. We can write $\partial_v \partial_r \mu_{AB}$ in terms of $R_{AB}$ (appendix \ref{app:elimination}) to deduce how it transforms. The result is that $\Delta (\partial_v \partial_r \mu_{AB})$ depends only on the first and second derivatives of $\log a$, and on $\beta_A$. By writing $\alpha$ in terms of $R_{vr}$ and $R_{AB}$ one sees that the same is true for\footnote{
An alternative way of obtaining these results is to observe that $\Delta (\partial_v \partial_r \mu_{AB})$ and $\Delta \alpha$ are of boost weight $0$ and dimension $2$ and are a sum of terms, each of which involves at least one factor of a (first or second) derivative of $\log a$. A derivative of $\log a$ has boost weight $0$ and dimension at least $1$, and so must multiply a boost weight $0$ term of dimension at most $1$. A term with boost weight $0$ that contains primitive factors with non-zero boost weight must have dimension at least $2$. So primitive factors of non-zero boost weight cannot appear in these quantities.
} $\Delta \alpha$. Thus $\Delta Z_i$ depends only on elements of $S_0$ and on the first and second derivatives of $\log a$. 

We can now rewrite $P_i$ and $N_i$ of \eqref{Deltas} in terms of our original basis, i.e., in terms of $S_{\pm 1}$ instead of $S_{\pm 1}'$. Recall that this does not spoil the property of having exactly one primitive factor of positive boost weight. This rewriting may generate extra factors (e.g. $\beta_A$) belonging to $S_0$. The result is that we have shown
\begin{equation}
 \label{Deltas2}
 \Delta s^{v}_{1}=\sum_i \hat{P}_i \hat{N}_i \hat{Z}_i
 \end{equation}
where $\hat{P}_i$ is an element of $S_2$ or $S_1$, with boost weight $b_i$,
$\hat{N}_i$ has boost weight $-b_i$ and is either an element of $S_{-2}$ or $S_{-1}$ or a product of two elements of $S_{-1}$, and $\hat{Z}_i$ depends only on elements of $S_0$ and on the first and second derivatives of $\log a$. 

Given an arbitrary dynamical black hole with metric $g_{\mu\nu}$, Ref. \cite{Hollands:2022} explains (section 3.3) how to construct a ``background'' black hole metric $\tilde{g}_{\mu\nu}$ (not necessarily satisfying any equations of motion) such that, on $C$, all background quantities of positive boost weight vanish whereas background quantities of non-positive boost weight agree with those of $g_{\mu\nu}$. Let $\delta g_{\mu\nu} = g_{\mu\nu}- \tilde{g}_{\mu\nu}$, for which all quantities of non-positive boost weight vanish on $C$. Consider the $1$-parameter family of metrics $g_{\mu\nu}(\lambda) = \tilde{g}_{\mu\nu} + \lambda \delta g_{\mu\nu}$, for which all quantities of non-positive boost weight agree with the corresponding quantities of $\tilde{g}_{\mu\nu}$ and $g_{\mu\nu}$ on $C$ and hence $\hat{N}_i[g(\lambda)]=\hat{N}_i[\tilde{g}]=N_i[g]$ and $\hat{Z}_i[g(\lambda)]=\hat{Z}_i[\tilde{g}]=Z_i[g]$. Since $\hat{P}_i$ is an element of $S_2$ or $S_1$ we have $\hat{P}_i[g(\lambda)] \propto \lambda$, i.e., the linear approximation to $\hat{P}_i[g(\lambda)]$ is exact: there are no terms of order $\lambda^2$ or higher. Hence we have $\hat{P}_i[g(\lambda)] \hat{N}_i[g(\lambda)]\hat{Z}_i[g(\lambda)] \propto \lambda$ so $\Delta s^v_1[g(\lambda)] \propto \lambda$. However, Ref. \cite{Hollands:2022} proved that the IWW entropy is gauge invariant to linear order in perturbations around any background solution defined as above, i.e., $\Delta s^v_1[g(\lambda)] =O(\lambda^2)$. Combining these results we have $\Delta s^v_1[g(\lambda)]=0$. Setting $\lambda=1$ we obtain $\Delta s^v_1[g]=0$ and we have proved that the IWW entropy is nonperturbatively gauge invariant for this class of theories.

\subsection{IWWHKR entropy}

Now we shall discuss gauge invariance of the IWWHKR entropy. In Section \ref{Classify}, we explained why the IWWHKR entropy density $S^v$ must be gauge invariant up to and including order $l^4$ terms, simply because there are no gauge-noninvariant terms that can appear at this order. This is confirmed by our calculations for the EGB and cubic Lagrangians. We shall now discuss the quantities calculated at order $l^6$ for the quartic Lagrangian above. As just discussed, the IWW part of the entropy is gauge invariant so we just need to discuss the transformation of the quantity $\varsigma_{6}^v$ given in equation \eqref{varsigma6}. This quantity is made out of  allowed terms, so we already know how all the terms transform. We again swap all $D$ derivatives for $\mathcal{D}$ derivatives on $K_{A B}$ and $\Bar{K}_{A B}$ terms. We then find 
\begin{equation}
    \begin{split}
        \varsigma^v_{6} =& \frac{k}{3} \bigg[ D_{(A}{\beta_{B)}} \Big(-48K K^{H I} \partial_{r}{\bar{K}_{H I}} \mu^{A B}+48K K^{H J} \bar{K}_{H}\,^{P} \bar{K}_{J P} \mu^{A B}-112K^{A B} K^{H I} \partial_{r}{\bar{K}_{H I}}+\\
        &112K^{A B} K^{H J} \bar{K}_{H}\,^{P} \bar{K}_{J P}\Big)+240D_{[A}{\beta_{B]}}K^{A I} K^{J P} \bar{K}^{B}\,_{I} \bar{K}_{J P}+\\
        &\beta_{A} \Big(-48K^{H I} \mathcal{D}^{A}{K} \partial_{r}{\bar{K}_{H I}}+48K^{H J} \mathcal{D}^{A}{K} \bar{K}_{H}\,^{P} \bar{K}_{J P}-48K \mathcal{D}^{A}{K^{H I}} \partial_{r}{\bar{K}_{H I}}+\\
        &48K \mathcal{D}^{A}{K^{H J}} \bar{K}_{H}\,^{P} \bar{K}_{J P}-32K^{G H} K^{J P} \bar{K}_{G H} \mathcal{D}^{A}{\bar{K}_{J P}}+32K^{G H} K^{I P} \bar{K}_{G H} \mathcal{D}_{I}{\bar{K}^{A}\,_{P}}-\\
        &32K^{A J} K^{G H} \bar{K}_{G H} \mathcal{D}^{P}{\bar{K}_{J P}}+32K^{A I} K^{G H} \bar{K}_{G H} \mathcal{D}_{I}{\bar{K}}-144K^{H I} \mathcal{D}^{G}{K^{A}\,_{G}} \partial_{r}{\bar{K}_{H I}}-\\
        &144K^{A G} K^{H I} \mathcal{D}_{G}\left(\partial_{r}{\bar{K}_{H I}}\right)+144K^{H J} \mathcal{D}^{G}{K^{A}\,_{G}} \bar{K}_{H}\,^{P} \bar{K}_{J P}+288K^{A I} K^{G J} \bar{K}_{G}\,^{P} \mathcal{D}_{I}{\bar{K}_{J P}}-\\
        &48K K^{H I} \mathcal{D}^{A}\left(\partial_{r}{\bar{K}_{H I}}\right)+96K K^{G J} \bar{K}_{G}\,^{P} \mathcal{D}^{A}{\bar{K}_{J P}}-144K^{A E} \mathcal{D}_{E}{K^{H I}} \partial_{r}{\bar{K}_{H I}}+\\
        &144K^{A E} \mathcal{D}_{E}{K^{H J}} \bar{K}_{H}\,^{P} \bar{K}_{J P}\Big)+\\
        &\beta_{A} \beta_{B} \left(16K^{A B} K^{H I} \partial_{r}{\bar{K}_{H I}}-16K^{A B} K^{H J} \bar{K}_{H}\,^{P} \bar{K}_{J P}\right) \bigg] + \text{homogeneous terms}
    \end{split}
\end{equation}
The $\beta_A$-dependence strongly suggests that this is not gauge invariant. To confirm this, we apply a gauge transformation and find
\begin{equation}
    \begin{split}
        \varsigma'^{v'}_{6} =& \varsigma^v_{6} + \frac{k}{3} \bigg[ 2D_{A}D_{B} \log a \Big(-48K K^{H I} \partial_{r}{\bar{K}_{H I}} \mu^{A B}+48K K^{H J} \bar{K}_{H}\,^{P} \bar{K}_{J P} \mu^{A B}-\\
        &112K^{A B} K^{H I} \partial_{r}{\bar{K}_{H I}}+112K^{A B} K^{H J} \bar{K}_{H}\,^{P} \bar{K}_{J P}\Big)+\\
        &2D_{A} \log a \Big(-48K^{H I} \mathcal{D}^{A}{K} \partial_{r}{\bar{K}_{H I}}+48K^{H J} \mathcal{D}^{A}{K} \bar{K}_{H}\,^{P} \bar{K}_{J P}-48K \mathcal{D}^{A}{K^{H I}} \partial_{r}{\bar{K}_{H I}}+\\
        &48K \mathcal{D}^{A}{K^{H J}} \bar{K}_{H}\,^{P} \bar{K}_{J P}-32K^{G H} K^{J P} \bar{K}_{G H} \mathcal{D}^{A}{\bar{K}_{J P}}+32K^{G H} K^{I P} \bar{K}_{G H} \mathcal{D}_{I}{\bar{K}^{A}\,_{P}}-\\
        &32K^{A J} K^{G H} \bar{K}_{G H} \mathcal{D}^{P}{\bar{K}_{J P}}+32K^{A I} K^{G H} \bar{K}_{G H} \mathcal{D}_{I}{\bar{K}}-144K^{H I} \mathcal{D}^{G}{K^{A}\,_{G}} \partial_{r}{\bar{K}_{H I}}-\\
        &144K^{A G} K^{H I} \mathcal{D}_{G}\left(\partial_{r}{\bar{K}_{H I}}\right)+144K^{H J} \mathcal{D}^{G}{K^{A}\,_{G}} \bar{K}_{H}\,^{P} \bar{K}_{J P}+288K^{A I} K^{G J} \bar{K}_{G}\,^{P} \mathcal{D}_{I}{\bar{K}_{J P}}-\\
        &48K K^{H I} \mathcal{D}^{A}\left(\partial_{r}{\bar{K}_{H I}}\right)+96K K^{G J} \bar{K}_{G}\,^{P} \mathcal{D}^{A}{\bar{K}_{J P}}-144K^{A E} \mathcal{D}_{E}{K^{H I}} \partial_{r}{\bar{K}_{H I}}+\\
        &144K^{A E} \mathcal{D}_{E}{K^{H J}} \bar{K}_{H}\,^{P} \bar{K}_{J P}\Big)+\\
        &4\left( \beta_{A} D_{B} \log a + D_A \log a D_B \log a \right) \left(16K^{A B} K^{H I} \partial_{r}{\bar{K}_{H I}}-16K^{A B} K^{H J} \bar{K}_{H}\,^{P} \bar{K}_{J P}\right) \bigg]
    \end{split}
\end{equation}
The IWWHKR entropy involves the above expression integrated over the horizon cross-section $C$. Integration by parts can be used to simplify the dependence on $a(x^A)$ in this integral:
\begin{equation}
    \begin{split}
        S_{IWWHKR}'=&S_{IWWHKR} + 4 \pi\int_C d^2 x \frac{kl^6}{3} \sqrt{\mu} \bigg[2 D_{A}{\log a} \Big(-32K^{C E} D^{B}{K^{A}\,_{B}} \partial_{r}{\bar{K}_{C E}}-\\
        &32K^{A B} D_{B}{K^{C E}} \partial_{r}{\bar{K}_{C E}}-32K^{A B} K^{C E} D_{B}\left(\partial_{r}{\bar{K}_{C E}}\right)+\\
        &32K^{C E} D^{B}{K^{A}\,_{B}} \bar{K}_{C}\,^{F} \bar{K}_{E F}+32K^{A B} D_{B}{K^{C E}} \bar{K}_{C}\,^{F} \bar{K}_{E F}+\\
        &64K^{A B} K^{C E} \bar{K}_{C}\,^{F} D_{B}{\bar{K}_{E F}}-32K^{B C} K^{E F} \bar{K}_{B C} D^{A}{\bar{K}_{E F}}-\\
        &16\beta^{A} K^{B C} K^{E F} \bar{K}_{B C} \bar{K}_{E F}+32K^{C E} K^{B F} \bar{K}_{C E} D_{B}{\bar{K}^{A}\,_{F}}+\\
        &16\beta^{B} K_{B}\,^{C} K^{E F} \bar{K}^{A}\,_{C} \bar{K}_{E F}-32K^{A C} K^{E F} \bar{K}_{E F} D^{B}{\bar{K}_{C B}}-\\
        &16\beta^{F} K^{A B} K^{C E} \bar{K}_{F B} \bar{K}_{C E}+32K^{A B} K^{C E} \bar{K}_{C E} D_{B}{\bar{K}}+\\
        &16\beta^{B} K^{A}\,_{B} K^{C E} \bar{K}_{C E} \bar{K}+32\beta^{B} K^{A}\,_{B} K^{C E} \partial_{r}{\bar{K}_{C E}}-\\
        &32\beta^{B} K^{A}\,_{B} K^{C E} \bar{K}_{C}\,^{F} \bar{K}_{E F}\Big)\\
        &+4D_{A}{\log a}D_{B}{\log a} \left(16K^{A B} K^{C E} \partial_{r}{\bar{K}_{C E}}-16 K^{A B} K^{C E} \bar{K}_{C}\,^{F} \bar{K}_{E F}\right) \bigg]
    \end{split}
\end{equation}
For gauge invariance to hold the coefficients of the terms linear and quadratic in $D_A \log a$ must vanish independently. However, these coefficients depend in a complicated way on expressions of quadratic order in positive boost weight quantities. There is no reason why they will vanish for a generic perturbation. Therefore the IWWHKR entropy of this $8$-derivative theory is not gauge invariant at order $l^6$. 

This statement concerns non-perturbative gauge invariance. However, since the IWWHKR satisfies the second law only to quadratic order in perturbation theory, modulo terms of order $l^8$, it is natural to demand gauge invariance in the same sense, i.e., only to quadratic order in perturbation theory, modulo terms of order $l^8$. Positive boost weight quantities are of at least linear order in perturbation theory so, to quadratic order, we can evaluate the negative boost weight quantities above in the unperturbed stationary black hole geometry. This might lead to extra cancellations. To investigate this, we shall focus on the $D_A \log a D_B \log a$ term above. It has coefficient proportional to $l^6 K^{AB} K^{CD} R_{rCrD}$ (using the expression for $R_{rCrD}$ in \cite{Hollands:2022}). To quadratic order, we can evaluate $R_{rCrD}$ in the stationary black hole geometry. Gauge invariance in the sense just discussed would require that $R_{rCrD}=O(l^2)$ on $C$. Using the equation of motion $R_{rr} = O(l^2)$, this gives $C_{rCrD} = O(l^2)$ on $C$, where the LHS is a component of the Weyl tensor. This is the statement that $n=\partial/\partial r$ is a principal null direction of the unperturbed black hole, modulo terms of order $l^2$. Recall that, by definition, $n$ is orthogonal to $C$. For a generic choice of $C$ there is no reason why $n$ should be close to being a principal null direction (although it can be in special cases e.g. a spherically symmetric cross-section of a spherically symmetric black hole). 
In particular, for a rotating black hole, one would expect an ``ingoing'' principal null direction to have non-zero rotation at the horizon (e.g. this is true for a Kerr black hole), whereas $n$ has vanishing rotation by definition. We conclude that, in general, the $D_A \log a D_B \log a$ term above is generically non-vanishing, of order $l^6$, to quadratic order in perturbation theory. Thus, for this $8$-derivative theory, the IWWHKR entropy is not gauge invariant in the desired sense. 

\section{Discussion}

\label{sec:discuss}

We have explained why the IWWHKR entropy is gauge invariant to order $l^4$, i.e., up to an including $6$-derivative terms in the Lagrangian. But we have seen that it is not gauge invariant at order $l^6$ for a specific $8$-derivative term in the Lagrangian. It is conceivable that if we allowed all possible $8$-derivative terms then demanding gauge invariance might lead to non-trivial relations between their coefficients, i.e., it might function as a selection rule for such theories. However we think this is unlikely, and that the lack of gauge invariance is a flaw of the HKR prescription. Nevertheless, we repeat that the IWWHKR entropy {\it is} gauge invariant for terms with up to $6$ derivatives in the Lagrangian, i.e., for the leading order EFT corrections to 4d vacuum gravity (and next to leading order for higher dimensional gravity). It is only for the next to leading order corrections that the problem arises so this is probably not a serious issue for practical applications. 

We shall end by mentioning possibilities for future work. HKR highlighted the issue of how the IWWHKR entropy transforms under EFT field redefinitions, and possible non-uniqueness of dynamical black hole entropy. This remains an interesting open question that we intend to return to. It would also be interesting to generalize the HKR algorithm beyond vacuum gravity, or gravity plus a scalar field, to include EFTs with more general matter content e.g. a Maxwell field. 

\subsection*{Acknowledgments}

We are grateful to Stefan Hollands, Shiraz Minwalla and Aron Wall for helpful discussions. ID is supported by an STFC studentship. HSR is supported by STFC grant no. ST/T000694/1.

\appendix

\section{Details of HKR Algorithm}

\subsection{Elimination Rules for Non-allowed Terms}

\label{app:elimination}

In our GNC expansion of $F\equiv E_{v v} - \partial_{v}\left[\frac{1}{\sqrt{\mu}} \partial_{v}\left(\sqrt{\mu} s^{v}_{IWW}\right) + D_{A}{ s^A }\right]$ on $\mathcal{N}$, we want to reduce the set of terms that appear up to $O(l^N)$ from the set given in (\ref{GeneralTerms}) to the set of "allowed terms" given in (\ref{AllowedTerms}). To eliminate non-allowed terms we study Ricci components and their covariant derivatives evaluated on $\mathcal{N}$. For the EGB, cubic and quartic Lagrangians above, we need the following:

\begin{equation}
    \begin{split}
        \partial_{v}{\beta_{A}} =& -2D^{B}{K_{A B}}+2D_{A}{K}-K \beta_{A}+2R_{v A}\\
        \partial_{r}{\beta_{A}} =& D^{B}{\bar{K}_{A B}}-D_{A}{\bar{K}}+\bar{K}_{A}\,^{B} \beta_{B} - \frac{1}{2}\bar{K} \beta_{A}-R_{r A}\\
        \partial_{v}{\bar{K}_{A B}} =& \frac{1}{2}R_{A B}[\mu]+K_{B}\,^{C} \bar{K}_{A C}+K_{A}\,^{C} \bar{K}_{B C} - \frac{1}{2}K \bar{K}_{A B} - \frac{1}{2}K_{A B} \bar{K} -\\
        &\frac{1}{4}D_{B}{\beta_{A}} - \frac{1}{4}\beta_{A} \beta_{B} - \frac{1}{4}D_{A}{\beta_{B}} - \frac{1}{2}R_{A B}\\
        \alpha =&  - \frac{1}{2}D^{A}{\beta_{A}}-\mu^{A B} \partial_{v}{\bar{K}_{A B}}+K^{A B} \bar{K}_{A B} - \frac{1}{2}\beta^{A} \beta_{A}-R_{r v}\\
        \end{split}
\end{equation}
\begin{equation}
    \begin{split}
        \partial_{r r}{\beta_{A}} =& \frac{2}{3}D^{B}\left(\partial_{r}{\bar{K}_{A B}}\right) - \frac{4}{3}D^{B}{\bar{K}_{A}\,^{C}} \bar{K}_{B C}+2D_{A}{\bar{K}^{B C}} \bar{K}_{B C}-2D^{B}{\bar{K}_{B}\,^{C}} \bar{K}_{A C}+\frac{4}{3}D^{B}{\bar{K}} \bar{K}_{A B} - \\
        &\frac{2}{3}D_{A}\left(\partial_{r}{\bar{K}_{B C}}\right) \mu^{B C} - \frac{2}{3}\bar{K} \partial_{r}{\beta_{A}}+\bar{K} \bar{K}_{A}\,^{B} \beta_{B} - \frac{2}{3}\beta_{A} \mu^{B C} \partial_{r}{\bar{K}_{B C}} - \frac{10}{3}\bar{K}_{A}\,^{B} \bar{K}_{B}\,^{C} \beta_{C}+\\
        &\frac{4}{3}\beta^{B} \partial_{r}{\bar{K}_{A B}}+2\bar{K}_{A}\,^{B} \partial_{r}{\beta_{B}}+\bar{K}^{B C} \bar{K}_{B C} \beta_{A}-\nabla_{r}{R_{A r}}\\
        \partial_{r v}{\bar{K}_{A B}} =& \frac{1}{2}D_{B}{D^{C}{\bar{K}_{A C}}} - \frac{1}{2}D_{B}{D_{A}{\bar{K}}} - \frac{1}{2}D^{C}{D_{C}{\bar{K}_{A B}}}+\frac{1}{2}D^{C}{D_{A}{\bar{K}_{B C}}}+\frac{3}{4}D^{C}{\beta_{A}} \bar{K}_{B C} -\\
        &\frac{1}{4}D_{B}{\beta_{A}} \bar{K}+\frac{1}{4}D_{B}{\beta^{C}} \bar{K}_{A C}+\frac{3}{4}D^{C}{\beta_{B}} \bar{K}_{A C}+2\bar{K}_{A}\,^{C} \partial_{v}{\bar{K}_{B C}}+K_{B}\,^{C} \partial_{r}{\bar{K}_{A C}}+\\
        &2\bar{K}_{B}\,^{C} \partial_{v}{\bar{K}_{A C}}+K_{A}\,^{C} \partial_{r}{\bar{K}_{B C}}+\frac{1}{4}D_{A}{\beta^{C}} \bar{K}_{B C} - \frac{1}{2}D^{C}{\beta_{C}} \bar{K}_{A B} - \frac{1}{2}\bar{K}_{A B} \mu^{C E} \partial_{v}{\bar{K}_{C E}} -\\
        &\frac{1}{2}K \partial_{r}{\bar{K}_{A B}} - \frac{1}{2}\bar{K} \partial_{v}{\bar{K}_{A B}} - \frac{1}{2}K_{A B} \mu^{C E} \partial_{r}{\bar{K}_{C E}} - \frac{1}{4}D_{A}{\beta_{B}} \bar{K}+\frac{1}{2}K_{B}\,^{C} \bar{K} \bar{K}_{A C}+\\
        &K \bar{K}_{A}\,^{C} \bar{K}_{B C}-2K^{C E} \bar{K}_{A C} \bar{K}_{B E}-3K_{B}\,^{C} \bar{K}_{A}\,^{E} \bar{K}_{C E}+\frac{1}{4}D^{C}{\bar{K}_{B C}} \beta_{A} - \frac{1}{4}D_{B}{\bar{K}} \beta_{A}+\\
        &\bar{K}_{B}\,^{C} \beta_{A} \beta_{C} - \frac{1}{4}\bar{K} \beta_{A} \beta_{B}-3K_{A}\,^{C} \bar{K}_{B}\,^{E} \bar{K}_{C E}+K^{C E} \bar{K}_{A B} \bar{K}_{C E}+K_{A B} \bar{K}^{C E} \bar{K}_{C E} -\\
        &\frac{3}{2}D^{C}{\bar{K}_{A B}} \beta_{C}+D_{B}{\bar{K}_{A}\,^{C}} \beta_{C}-\bar{K}_{A B} \beta^{C} \beta_{C}+\bar{K}_{A}\,^{C} \beta_{B} \beta_{C} - \frac{1}{2}\bar{K}_{B}\,^{C} R_{A C}[\mu]+\\
        &\frac{1}{2}K_{A}\,^{C} \bar{K} \bar{K}_{B C}+\frac{1}{4}D^{C}{\bar{K}_{A C}} \beta_{B} - \frac{1}{4}D_{A}{\bar{K}} \beta_{B} - \frac{1}{2}\bar{K}^{C E} R_{A C B E}[\mu]+D_{A}{\bar{K}_{B}\,^{C}} \beta_{C} -\\
        &\frac{1}{2}D_{B}\left(\partial_{r}{\beta_{A}}\right)-\bar{K}_{A B} \alpha - \frac{3}{4}\beta_{A} \partial_{r}{\beta_{B}} - \frac{3}{4}\beta_{B} \partial_{r}{\beta_{A}} - \frac{1}{2}D_{A}\left(\partial_{r}{\beta_{B}}\right) - \frac{1}{2}\nabla_{r}{R_{A B}}\\
        \partial_{r}{\alpha} =& \frac{1}{3}D^{A}{\beta^{B}} \bar{K}_{A B} - \frac{1}{4}\bar{K} \beta^{A} \beta_{A} - \frac{1}{3}D^{A}\left(\partial_{r}{\beta_{A}}\right)+\frac{1}{2}D^{A}{\bar{K}_{A}\,^{B}} \beta_{B} - \frac{1}{3}D^{A}{\bar{K}} \beta_{A} - \frac{1}{3}\bar{K} \alpha -\\
        &\frac{1}{3}\mu^{A B} \partial_{r v}{\bar{K}_{A B}}+\bar{K}^{A B} \partial_{v}{\bar{K}_{A B}}+\frac{1}{3}K^{A B} \partial_{r}{\bar{K}_{A B}} - \frac{4}{3}K^{A B} \bar{K}_{A}\,^{C} \bar{K}_{B C}+\frac{5}{6}\bar{K}^{A B} \beta_{A} \beta_{B} -\\
        &\frac{7}{6}\beta^{A} \partial_{r}{\beta_{A}}+\frac{1}{3}\nabla_{r}{R_{v r}}
    \end{split}
\end{equation}
We also need elimination rules for $\partial_{v v}{\beta_{A}}, \partial_{v r}{\beta_{A}}, \partial_{v v v}{\beta_{A}}, \partial_{v v r}{\beta_{A}}, \partial_{v v}{\Bar{K}_{A B}}$ and $\partial_{v v}{\alpha}$, but these can be found by taking $\partial_{v}$ derivatives of the above. We can then use the equation of motion $R_{\mu\nu} = \frac{2}{d-2} \Lambda g_{\mu \nu} - \frac{1}{d-2} g^{\rho \sigma} H_{\rho \sigma} g_{\mu \nu} + H_{\mu \nu} + O(l^N)$ to exchange $R_{\mu \nu}$ for $\frac{2}{d-2} \Lambda g_{\mu \nu}$ plus terms that are at least $O(l^2)$. We can repeat this process until all non-allowed terms are of order $l^N$, at which point we can neglect them for our calculation of the HKR entropy.

\subsection{Calculation of $\varsigma_n^{v}$ Terms}

\label{app:aj}

The HKR algorithm involves finding numbers $a_j$ such that

\begin{equation} \label{dKDdK}
    (\partial_{v}^{p_1} K) \, (D^{k}{\partial_{v}^{p_2}K}) A_{k, p_1, p_2} = \partial_{v}\left[\frac{1}{\sqrt{\mu}} \partial_{v}\left(\sqrt{\mu} \sum_{j=1}^{p_1+p_2-1} a_j (\partial_{v}^{p_1+p_2-1-j} K) \, (D^{k}{\partial_{v}^{j-1}K }) A_{k, p_1, p_2} \right)\right] + ...
\end{equation}
where the ellipsis denotes terms of the form $(\partial_{v}^{\Bar{p}_1} K) \, (D^{\bar{k}}{\partial_{v}^{\Bar{p}_2}K}) A_{\Bar{k}, \Bar{p}_1, \Bar{p}_2}$ with $\Bar{p}_1+\Bar{p}_2 < p_1+p_2$ or $\Bar{p}_1=0$ or $\Bar{p}_2=0$. 

How to calculate the $a_j$ for any $k, p_1, p_2$ is given in \cite{Hollands:2022}, and is as follows. When the derivative term on the RHS is expanded out we get a set of $p_1 + p_2 -1$ linear equations on the $a_j$ in order to satisfy the required conditions. The linear equations can be written in matrix form as 
\begin{equation}
    M_{p_1 + p_2 -1} \textbf{a} = \textbf{v}_{p_2}
\end{equation}
where
\begin{equation}
    M_{p_1 + p_2 -1} = \begin{pmatrix}
        2 & 1 & 0 & 0 & ... & 0 \\
        1 & 2 & 1 & 0 & ... & 0 \\
        0 & 1 & 2 & 1 & ... & 0 \\
        ... & ... & ... & ... & ... & ... \\
        0 & ... & ... & 0 & 1 & 2 
    \end{pmatrix}, \quad
    \textbf{a}= \begin{pmatrix}
        a_1\\
        a_2\\
        ...\\
        a_{p_1 + p_2 -1}
    \end{pmatrix}, \quad
    \textbf{v}_{p_2} = \begin{pmatrix}
        0\\
        ...\\
        0\\
        1\\
        0\\
        ...\\
        0
    \end{pmatrix}
\end{equation}
where $(\textbf{v}_{p_2})_{p_2} = 1$ and $(\textbf{v}_{p_2})_{j} = 0$ for $j\neq p_2$. $M_{p_1+p_2-1}$ can be shown to have non-vanishing determinant, and so the system of equations has a unique solution. 

When performing the HKR algorithm for specific Lagrangians in practice, one should investigate for which values of $k, p_1, p_2$ do terms of the form on the LHS of (\ref{dKDdK}) appear in $F$, and pre-calculate the corresponding RHS of (\ref{dKDdK}). For the EGB, cubic and quartic Lagrangians above, the only relevant terms that appear are $\partial_{v} K_{A B} \partial_{v} K_{C D}$, $\partial_{v} K_{A B} D_{C} \partial_{v} K_{D E}$, $\partial_{v} K_{A B} D_{C} D_{D}\partial_{v} K_{E F}$ which have $(p_1, p_2)=(1,1)$ and $a_1=1$, and $\partial_{v} K_{A B} \partial_{v v} K_{C D}$ which has $(p_1, p_2)=(1,2)$ and $(a_1, a_2)= (-\frac{1}{3}, \frac{2}{3})$.

\section{Results for $s^A$}

\subsection{$s^{A}$ for Cubic Riemann Lagrangians}

\label{app:sAcubic}

\begin{equation}
    s^{A} = l^4( k_{1} s^{A}_{even} + k_{2} s^{A}_{odd} )
\end{equation}
where
\begin{equation}
    \begin{split}
   s^{A}_{even} =& -18D^{B}{K^{A C}} D_{B}{\beta_{C}}+12D^{B}{K^{A C}} D_{C}{\beta_{B}}+9D_{B}{\beta_{C}} K^{A C} \beta^{B}-6D_{B}{\beta_{C}} K^{A B} \beta^{C}+\\
   &12D^{B}{K^{A}\,_{C}} K^{C E} \bar{K}_{B E}-12D_{B}{K^{A C}} K^{B E} \bar{K}_{C E}-12K^{A B} K_{B}\,^{C} \bar{K}_{C E} \beta^{E}+\\
   &6K^{A B} K^{C E} \bar{K}_{B E} \beta_{C}-24\mu^{A B} \mu^{C E} \partial_{r}{\beta_{C}} \partial_{v}{K_{B E}}+12K^{A B} K_{B}\,^{C} \partial_{r}{\beta_{C}}+\\
   &12\bar{K}^{B}\,_{C} \beta^{C} \mu^{A E} \partial_{v}{K_{B E}}-6D^{A}{\beta^{B}} \partial_{v}{\beta_{B}}-12\mu^{A B} \mu^{C E} \partial_{v}{\bar{K}_{B C}} \partial_{v}{\beta_{E}}+\\
   &6K^{B C} \bar{K}^{A}\,_{C} \partial_{v}{\beta_{B}}+6K^{B C} K_{B}\,^{E} \bar{K}^{A}\,_{E} \beta_{C}-3\beta^{A} \beta^{B} \partial_{v}{\beta_{B}}-12\alpha \mu^{A B} \partial_{v}{\beta_{B}}-\\
   &12K^{A B} \alpha \beta_{B}-3\beta^{B} \beta_{B} \mu^{A C} \partial_{v}{\beta_{C}}-6D^{B}{K_{B}\,^{C}} D^{A}{\beta_{C}}-12D^{B}{K_{B}\,^{C}} \mu^{A E} \partial_{v}{\bar{K}_{C E}}-\\
   &3D^{B}{K_{B}\,^{C}} \beta^{A} \beta_{C}-6D^{B}{D_{B}{\beta_{C}}} K^{A C}-12D^{B}\left(\partial_{v}{\bar{K}_{B C}}\right) K^{A C}-3D^{B}{\beta_{B}} K^{A C} \beta_{C}-\\
   &12D^{B}{K^{A C}} \partial_{v}{\bar{K}_{B C}}-3D^{B}{K^{A C}} \beta_{B} \beta_{C}+12K^{A B} \beta^{C} \partial_{v}{\bar{K}_{B C}}-6D_{B}{D^{A}{\beta_{C}}} K^{B C}-\\
   &12D_{B}\left(\partial_{v}{\bar{K}_{C E}}\right) K^{B C} \mu^{A E}-3D_{B}{\beta_{C}} K^{B C} \beta^{A}-3D_{B}{\beta^{A}} K^{C B} \beta_{C}
   \end{split}
\end{equation}
\begin{equation}
    \begin{split}
    s^{A}_{odd} =& -8D_{B}{\beta_{C}} \epsilon^{B C} \mu^{A E} \partial_{v}{\beta_{E}}-12D_{B}{\beta_{C}} K^{A E} \beta_{E} \epsilon^{B C}-8K_{B}\,^{C} \bar{K}_{C E} \epsilon^{B E} \mu^{A F} \partial_{v}{\beta_{F}}-\\
   &8K_{B}\,^{C} K^{A E} \bar{K}_{C F} \beta_{E} \epsilon^{B F}-16\epsilon^{B C} \mu^{A E} \partial_{r}{\beta_{B}} \partial_{v}{K_{C E}}-8K_{B}\,^{C} K^{A}\,_{C} \epsilon^{B E} \partial_{r}{\beta_{E}}+\\
   &8\bar{K}_{B C} \beta^{B} \epsilon^{C E} \mu^{A F} \partial_{v}{K_{E F}}+ 8K_{B}\,^{C} K^{A}\,_{C} \bar{K}_{E F} \beta^{E} \epsilon^{B F}-4D^{A}{\beta_{B}} \epsilon^{B C} \partial_{v}{\beta_{C}}-\\
   &8\epsilon^{B C} \mu^{A E} \partial_{v}{\bar{K}_{B E}} \partial_{v}{\beta_{C}} + 4K_{B}\,^{C} \bar{K}^{A}\,_{C} \epsilon^{B E} \partial_{v}{\beta_{E}}+4K_{B}\,^{C} K_{E}\,^{F} \bar{K}^{A}\,_{C} \beta_{F} \epsilon^{B E}-\\
   &2\beta_{B} \beta^{A} \epsilon^{B C} \partial_{v}{\beta_{C}} + 16D_{B}{K_{C}\,^{A}} \alpha \epsilon^{B C}+8K_{B}\,^{A} \alpha \beta_{C} \epsilon^{B C}+ 4D_{B}{K_{C}\,^{A}} \beta^{E} \beta_{E} \epsilon^{B C} -\\
   &12D^{B}{K_{C}\,^{E}} D_{B}{\beta_{E}} \epsilon^{A C}+8D^{B}{K_{C}\,^{E}} D_{E}{\beta_{B}} \epsilon^{A C}+6D_{B}{\beta_{C}} K_{E}\,^{C} \beta^{B} \epsilon^{A E}-\\
   &4D_{B}{\beta_{C}} K_{E}\,^{B} \beta^{C} \epsilon^{A E}+8D^{B}{K_{C E}} K^{C F} \bar{K}_{B F} \epsilon^{A E}-8D_{B}{K_{C}\,^{E}} K^{B F} \bar{K}_{E F} \epsilon^{A C}-\\
   &8K_{B}\,^{C} K_{C}\,^{E} \bar{K}_{E F} \beta^{F} \epsilon^{A B}+4K_{B}\,^{C} K^{E F} \bar{K}_{C F} \beta_{E} \epsilon^{A B}-16\epsilon^{A B} \mu^{C E} \partial_{r}{\beta_{C}} \partial_{v}{K_{B E}}+\\
   &8K_{B}\,^{C} K_{C}\,^{E} \epsilon^{A B} \partial_{r}{\beta_{E}}+8\bar{K}^{B}\,_{C} \beta^{C} \epsilon^{A E} \partial_{v}{K_{B E}}-4D_{B}{\beta^{C}} \epsilon^{A B} \partial_{v}{\beta_{C}}-\\
   &8\epsilon^{A B} \mu^{C E} \partial_{v}{\bar{K}_{B C}} \partial_{v}{\beta_{E}}+4K^{B C} \bar{K}_{C E} \epsilon^{A E} \partial_{v}{\beta_{B}}+4K^{B C} K_{B}\,^{E} \bar{K}_{E F} \beta_{C} \epsilon^{A F}-\\
   &2\beta_{B} \beta^{C} \epsilon^{A B} \partial_{v}{\beta_{C}}-8\alpha \epsilon^{A B} \partial_{v}{\beta_{B}}-8K_{B}\,^{C} \alpha \beta_{C} \epsilon^{A B}-2\beta^{B} \beta_{B} \epsilon^{A C} \partial_{v}{\beta_{C}}-\\
   &8D_{B}{K_{C}\,^{A}} R[\mu] \epsilon^{B C}-4K_{B}\,^{A} R[\mu] \beta_{C} \epsilon^{B C}-8D^{B}{K^{A}\,_{C}} K_{E}\,^{C} \bar{K}_{B F} \epsilon^{E F}+\\
   &8D_{B}{K^{A C}} K_{E}\,^{B} \bar{K}_{C F} \epsilon^{E F}-4K_{B}\,^{C} K^{A E} \bar{K}_{E F} \beta_{C} \epsilon^{B F}+16D_{B}{\bar{K}_{C}\,^{E}} \epsilon^{B C} \mu^{A F} \partial_{v}{K_{E F}}-\\
   &8D_{B}{\bar{K}_{C E}} K^{A F} K_{F}\,^{C} \epsilon^{B E}-8\bar{K}_{B}\,^{C} \beta_{E} \epsilon^{B E} \mu^{A F} \partial_{v}{K_{C F}}+4K^{A B} K_{B}\,^{C} \bar{K}_{C E} \beta_{F} \epsilon^{E F}+\\
   &4D_{B}{K_{C}\,^{E}} D^{A}{\beta_{E}} \epsilon^{B C}+8D_{B}{K_{C}\,^{E}} \epsilon^{B C} \mu^{A F} \partial_{v}{\bar{K}_{E F}}-8D_{B}{K_{C E}} K^{C F} \bar{K}^{A}\,_{F} \epsilon^{B E}-\\
   &4K_{B}\,^{C} K_{C}\,^{E} \bar{K}^{A}\,_{E} \beta_{F} \epsilon^{B F}+2D_{B}{K_{C}\,^{E}} \beta^{A} \beta_{E} \epsilon^{B C}+4D^{B}{K_{B C}} D^{A}{\beta_{E}} \epsilon^{C E}+\\
   &8D^{B}{K_{B C}} \epsilon^{C E} \mu^{A F} \partial_{v}{\bar{K}_{E F}}+2D^{B}{K_{B C}} \beta_{E} \beta^{A} \epsilon^{C E}-4D^{B}{D_{B}{\beta_{C}}} K_{E}\,^{A} \epsilon^{C E}-\\
   &8D^{B}\left(\partial_{v}{\bar{K}_{B C}}\right) K_{E}\,^{A} \epsilon^{C E}+2D^{B}{\beta_{B}} K_{C}\,^{A} \beta_{E} \epsilon^{C E}+2D_{B}{\beta_{C}} K_{E}\,^{A} \beta^{C} \epsilon^{B E}+\\
   &4D^{B}{K_{C}\,^{A}} D_{B}{\beta_{E}} \epsilon^{C E}+8D^{B}{K_{C}\,^{A}} \epsilon^{C E} \partial_{v}{\bar{K}_{B E}}+2D^{B}{K_{C}\,^{A}} \beta_{B} \beta_{E} \epsilon^{C E}-\\
   &8K_{B}\,^{A} \beta^{C} \epsilon^{B E} \partial_{v}{\bar{K}_{C E}}-4D_{B}{D^{A}{\beta_{C}}} K_{E}\,^{B} \epsilon^{C E}-8D_{B}\left(\partial_{v}{\bar{K}_{C E}}\right) K_{F}\,^{B} \epsilon^{C F} \mu^{A E}-\\
   &2D_{B}{\beta_{C}} K_{E}\,^{B} \beta^{A} \epsilon^{C E}+2D_{B}{\beta^{A}} K_{C}\,^{B} \beta_{E} \epsilon^{C E}-4D_{B}{D_{C}{\beta_{E}}} K^{A E} \epsilon^{B C}-\\
   &6D_{B}{\beta_{C}} K^{A C} \beta_{E} \epsilon^{B E}-4D_{B}{K^{A C}} D_{E}{\beta_{C}} \epsilon^{B E}-8D_{B}{K^{A C}} \epsilon^{B E} \partial_{v}{\bar{K}_{C E}}-\\
   &2D_{B}{K^{A C}} \beta_{C} \beta_{E} \epsilon^{B E}+8K^{A B} \beta_{C} \epsilon^{C E} \partial_{v}{\bar{K}_{B E}}-4D_{B}{D^{A}{\beta_{C}}} K_{E}\,^{C} \epsilon^{B E}-\\
   &8D_{B}\left(\partial_{v}{\bar{K}_{C E}}\right) K_{F}\,^{C} \epsilon^{B F} \mu^{A E}-2D_{B}{\beta_{C}} K_{E}\,^{C} \beta^{A} \epsilon^{B E}-2D_{B}{\beta^{A}} K_{C}\,^{E} \beta_{E} \epsilon^{B C}-\\
   &8D^{B}{K_{C}\,^{E}} \epsilon^{A C} \partial_{v}{\bar{K}_{B E}}-2D^{B}{K_{C}\,^{E}} \beta_{B} \beta_{E} \epsilon^{A C}+8K_{B}\,^{C} \beta^{E} \epsilon^{A B} \partial_{v}{\bar{K}_{C E}}-\\
   &4D_{B}{D_{C}{\beta_{E}}} K^{B E} \epsilon^{A C}-8D_{B}\left(\partial_{v}{\bar{K}_{C E}}\right) K^{B C} \epsilon^{A E}-2D_{B}{\beta_{C}} K^{B C} \beta_{E} \epsilon^{A E}-\\
   &2D_{B}{\beta_{C}} K^{E B} \beta_{E} \epsilon^{A C}-4D^{B}{K_{B}\,^{C}} D_{E}{\beta_{C}} \epsilon^{A E}-8D^{B}{K_{B}\,^{C}} \epsilon^{A E} \partial_{v}{\bar{K}_{C E}}-\\
   &2D^{B}{K_{B}\,^{C}} \beta_{C} \beta_{E} \epsilon^{A E}-4D^{B}{D_{B}{\beta_{C}}} K_{E}\,^{C} \epsilon^{A E}-8D^{B}\left(\partial_{v}{\bar{K}_{B C}}\right) K_{E}\,^{C} \epsilon^{A E}-\\
   &2D^{B}{\beta_{B}} K_{C}\,^{E} \beta_{E} \epsilon^{A C}-8D_{B}\left(\partial_{v}{\bar{K}_{C E}}\right) K^{A C} \epsilon^{B E}
    \end{split}
\end{equation}

\subsection{$s^{A}_{6}$ for Quartic Riemann Lagrangian}

\label{app:s6A}

\begin{equation}
    \begin{split}
        s^{A}_{6} =& k l^6 \Big[ 8\mu^{A B} \partial_{v}{\beta_{B}} {R[\mu]}^{2}-16K \bar{K} R[\mu] \mu^{A E} \partial_{v}{\beta_{E}}+16K^{B C} \bar{K}_{B C} R[\mu] \mu^{A E} \partial_{v}{\beta_{E}}-\\
        &16K^{A B} K \bar{K} R[\mu] \beta_{B}+16K^{A B} K^{C E} \bar{K}_{C E} R[\mu] \beta_{B}+16K^{B C} K_{B C} \bar{K}^{E F} \bar{K}_{E F} \mu^{A G} \partial_{v}{\beta_{G}}+\\
        &16K^{A B} K^{C E} K_{C E} \bar{K}^{F G} \bar{K}_{F G} \beta_{B}+16K^{B C} K^{E F} \bar{K}_{B C} \bar{K}_{E F} \mu^{A G} \partial_{v}{\beta_{G}}+\\
        &16K^{A B} K^{C E} K^{F G} \bar{K}_{C E} \bar{K}_{F G} \beta_{B}-32K^{B C} K_{B}\,^{E} \bar{K}_{C}\,^{F} \bar{K}_{E F} \mu^{A G} \partial_{v}{\beta_{G}}-\\
        &32K^{A B} K^{C E} K_{C}\,^{F} \bar{K}_{E}\,^{G} \bar{K}_{F G} \beta_{B}+64D^{B}{K^{C E}} D_{B}{\bar{K}_{C E}} \mu^{A F} \partial_{v}{\beta_{F}}-\\
        &64D^{B}{K^{C E}} D_{C}{\bar{K}_{B E}} \mu^{A F} \partial_{v}{\beta_{F}}-32D_{B}{\bar{K}_{C E}} K^{C E} \beta^{B} \mu^{A F} \partial_{v}{\beta_{F}}+\\
        &32D_{B}{\bar{K}_{C E}} K^{B C} \beta^{E} \mu^{A F} \partial_{v}{\beta_{F}}+64D^{B}{K^{C E}} D_{B}{\bar{K}_{C E}} K^{A F} \beta_{F}-64D^{B}{K^{C E}} D_{C}{\bar{K}_{B E}} K^{A F} \beta_{F}-\\
        &32D_{B}{\bar{K}_{C E}} K^{A F} K^{C E} \beta_{F} \beta^{B}+32D_{B}{\bar{K}_{C E}} K^{A F} K^{B C} \beta_{F} \beta^{E}+32D^{B}{K^{C E}} \bar{K}_{C E} \beta_{B} \mu^{A F} \partial_{v}{\beta_{F}}-\\
        &32D^{B}{K^{C E}} \bar{K}_{B E} \beta_{C} \mu^{A F} \partial_{v}{\beta_{F}}-16K^{B C} \bar{K}_{B C} \beta^{E} \beta_{E} \mu^{A F} \partial_{v}{\beta_{F}}+16K^{B C} \bar{K}_{C E} \beta_{B} \beta^{E} \mu^{A F} \partial_{v}{\beta_{F}}+\\
        &32D^{B}{K^{C E}} K^{A F} \bar{K}_{C E} \beta_{B} \beta_{F}-32D^{B}{K^{C E}} K^{A F} \bar{K}_{B E} \beta_{C} \beta_{F}-16K^{A B} K^{C E} \bar{K}_{C E} \beta_{B} \beta^{F} \beta_{F}+\\
        &16K^{A B} K^{C E} \bar{K}_{E F} \beta_{B} \beta_{C} \beta^{F}+32\mu^{A B} \mu^{C E} \mu^{F G} \partial_{r}{\bar{K}_{C F}} \partial_{v}{K_{E G}} \partial_{v}{\beta_{B}}-\\
        &32K^{B C} K_{B}\,^{E} \mu^{A F} \partial_{r}{\bar{K}_{C E}} \partial_{v}{\beta_{F}}+32K^{A B} \beta_{B} \mu^{C E} \mu^{F G} \partial_{r}{\bar{K}_{C F}} \partial_{v}{K_{E G}}-\\
        &32K^{A B} K^{C E} K_{C}\,^{F} \beta_{B} \partial_{r}{\bar{K}_{E F}}-32\bar{K}^{B C} \bar{K}^{E}\,_{C} \mu^{A F} \partial_{v}{K_{B E}} \partial_{v}{\beta_{F}}-32K^{A B} \bar{K}^{C E} \bar{K}^{F}\,_{E} \beta_{B} \partial_{v}{K_{C F}}-\\
        &16D^{B}{\beta^{C}} D_{B}{\beta_{C}} \mu^{A E} \partial_{v}{\beta_{E}}+32D^{B}{\beta^{C}} D_{C}{\beta_{B}} \mu^{A E} \partial_{v}{\beta_{E}}+32D^{B}{\beta_{C}} K^{C E} \bar{K}_{B E} \mu^{A F} \partial_{v}{\beta_{F}}+\\
        &32D^{B}{\beta_{C}} K^{A E} K^{C F} \bar{K}_{B F} \beta_{E}-64D_{B}{\beta^{C}} K^{B E} \bar{K}_{C E} \mu^{A F} \partial_{v}{\beta_{F}}-64D_{B}{\beta^{C}} K^{A E} K^{B F} \bar{K}_{C F} \beta_{E}+\\
        &64K^{B C} K^{E F} \bar{K}_{B E} \bar{K}_{C F} \mu^{A G} \partial_{v}{\beta_{G}}+64K^{A B} K^{C E} K^{F G} \bar{K}_{C F} \bar{K}_{E G} \beta_{B}+64D^{B}{\beta^{C}} \mu^{A E} \partial_{v}{\bar{K}_{B C}} \partial_{v}{\beta_{E}}+\\
        &16D_{B}{\beta_{C}} \beta^{B} \beta^{C} \mu^{A E} \partial_{v}{\beta_{E}}+64\mu^{A B} \mu^{C E} \mu^{F G} \partial_{v}{\bar{K}_{C F}} \partial_{v}{\bar{K}_{E G}} \partial_{v}{\beta_{B}}-64K^{B C} \bar{K}_{B}\,^{E} \mu^{A F} \partial_{v}{\bar{K}_{C E}} \partial_{v}{\beta_{F}}-\\
        &64K^{A B} K^{C E} \bar{K}_{C}\,^{F} \beta_{B} \partial_{v}{\bar{K}_{E F}}+32\beta^{B} \beta^{C} \mu^{A E} \partial_{v}{\bar{K}_{B C}} \partial_{v}{\beta_{E}}+6\beta^{B} \beta_{B} \beta^{C} \beta_{C} \mu^{A E} \partial_{v}{\beta_{E}}-\\
        &32\mu^{A B} \mu^{C E} \partial_{r}{\beta_{C}} \partial_{v}{\beta_{B}} \partial_{v}{\beta_{E}}-32K^{B C} \beta_{B} \mu^{A E} \partial_{r}{\beta_{C}} \partial_{v}{\beta_{E}}-32K^{A B} \beta_{B} \mu^{C E} \partial_{r}{\beta_{C}} \partial_{v}{\beta_{E}}-\\
        &32K^{A B} K^{C E} \beta_{B} \beta_{C} \partial_{r}{\beta_{E}}+16\bar{K}_{B}\,^{C} \beta^{B} \mu^{A E} \partial_{v}{\beta_{C}} \partial_{v}{\beta_{E}}+16K^{A B} \bar{K}_{C}\,^{E} \beta_{B} \beta^{C} \partial_{v}{\beta_{E}}+\\
        &32\mu^{A B} \partial_{v}{\beta_{B}} {\alpha}^{2}+16\alpha \beta^{B} \beta_{B} \mu^{A C} \partial_{v}{\beta_{C}}+8D^{B}{K^{A}\,_{B}} {R[\mu]}^{2}-16D^{B}{K^{A}\,_{B}} D^{C}{\beta^{E}} D_{C}{\beta_{E}}+\\
        &32D^{B}{K^{A}\,_{B}} D^{C}{\beta^{E}} D_{E}{\beta_{C}}+64D^{B}{K^{A}\,_{B}} D^{C}{\beta^{E}} \partial_{v}{\bar{K}_{C E}}+16D^{B}{K^{A}\,_{B}} D_{C}{\beta_{E}} \beta^{C} \beta^{E}+\\
        &64D^{B}{K^{A}\,_{B}} \mu^{C E} \mu^{F G} \partial_{v}{\bar{K}_{C F}} \partial_{v}{\bar{K}_{E G}}+32D^{B}{K^{A}\,_{B}} \beta^{C} \beta^{E} \partial_{v}{\bar{K}_{C E}}+6D^{B}{K^{A}\,_{B}} \beta^{C} \beta_{C} \beta^{E} \beta_{E}+\\
        &32D^{B}{K^{A}\,_{B}} {\alpha}^{2}+16D^{B}{K^{A}\,_{B}} \alpha \beta^{C} \beta_{C}+16D_{B}{R[\mu]} K^{A B} R[\mu]-32D_{B}{D_{C}{\beta_{E}}} D^{C}{\beta^{E}} K^{A B}+\\
        &64D_{B}{D_{C}{\beta_{E}}} D^{E}{\beta^{C}} K^{A B}+64D_{B}{D^{C}{\beta^{E}}} K^{A B} \partial_{v}{\bar{K}_{C E}}+16D_{B}{D_{C}{\beta_{E}}} K^{A B} \beta^{C} \beta^{E}+\\
        &64D^{B}{\beta^{C}} D_{E}\left(\partial_{v}{\bar{K}_{B C}}\right) K^{A E}+128D_{B}\left(\partial_{v}{\bar{K}_{C E}}\right) K^{A B} \mu^{C F} \mu^{E G} \partial_{v}{\bar{K}_{F G}}+\\
        &32D_{B}\left(\partial_{v}{\bar{K}_{C E}}\right) K^{A B} \beta^{C} \beta^{E}+16D_{B}{\beta^{C}} D_{C}{\beta_{E}} K^{A B} \beta^{E}+64D_{B}{\beta^{C}} K^{A B} \beta^{E} \partial_{v}{\bar{K}_{C E}}+\\
        &24D_{B}{\beta_{C}} K^{A B} \beta^{C} \beta_{E} \beta^{E}+16D_{B}{\beta^{C}} D_{E}{\beta_{C}} K^{A B} \beta^{E}+64D_{B}{\alpha} K^{A B} \alpha+\\
        &16D_{B}{\alpha} K^{A B} \beta^{C} \beta_{C}+32D_{B}{\beta_{C}} K^{A B} \alpha \beta^{C} \Big]
    \end{split}
\end{equation}

\end{document}